\newif\ifsingle
\singletrue 

\newif\ifFullVersion
  \FullVersiontrue 

\ifsingle
\documentclass[12pt,draftclsnofoot, onecolumn]{IEEEtran}		
\else		
\documentclass[10pt,final, twocolumn]{IEEEtran}
\fi

\usepackage{times}
\usepackage{amsmath,dsfont}
\usepackage{amssymb,amsthm}
\usepackage{epsfig,verbatim}
\usepackage{subfigure}
\usepackage{setspace}
\usepackage{color}
\usepackage{cite}
\usepackage{epstopdf}
\usepackage{graphics}
\usepackage{graphicx}
\usepackage{accents}
\usepackage{acronym}
\usepackage[bookmarks,colorlinks]{hyperref}
\usepackage{booktabs}
\usepackage{mathtools}
\usepackage{algorithm}
\usepackage{algorithmic}
\usepackage{enumitem}

\newcommand{\myVec}[1]{{\boldsymbol{#1}}}
\newcommand{\myMat}[1]{{\boldsymbol{#1}}}

\newcommand{\E}{\mathds{E}}		 			

\newtheorem{theorem}{Theorem}

\newtheorem{proposition}{Proposition}
\newtheorem{lemma}{Lemma}

\ifsingle

\else

\fi 

\acrodef{adc}[ADC]{Analog-to-Digital Convertor}
\acrodef{dac}[DAC]{digital-to-analog convertor}
\acrodef{cs}[CS]{Compressed Sensing}
\acrodef{dtft}[DTFT]{discrete-time Fourier transform}
\acrodef{dnn}[DNN]{deep neural network} 
\acrodef{csi}[CSI]{Channel State Information}
\acrodef{map}[MAP]{maximum a-posteriori probability}
\acrodef{snr}[SNR]{Signal-to-Noise Ratio}
\acrodef{sinr}[SINR]{signal-to-interference-and-noise ratio}
\acrodef{bs}[BS]{Base Station} 
\acrodef{iot}[IOT]{Internet of Things}
\acrodef{mimo}[MIMO]{Multiple-Input Multiple-Output}
\acrodef{mse}[MSE]{Mean-Squared Error}
\acrodef{pdf}[PDF]{probability density function}
\acrodef{rv}[RV]{random variable}
\acrodef{tdd}[TDD]{time division duplexing}
\acrodef{rs}[RS]{Reed-Solomon}
\acrodef{lti}[LTI]{linear time-invariant}
\acrodef{wss}[WSS]{wide-sense stationary}
\acrodef{psd}[PSD]{power spectral density}
\acrodef{ser}[SER]{symbol error rate} 
\acrodef{ber}[BER]{bit error rate} 
\acrodef{isi}[ISI]{intersymbol interference}  
\acrodef{awgn}[AWGN]{Additive White Gaussian Noise} 
\acrodef{ut}[UT]{User Terminal} 
\acrodef{mmw}[mmWave]{millimeter wave}
\acrodef{ris}[RIS]{Reconfigurable Intelligent Surface} 
\acrodef{dma}[DMA]{Dynamic Metasurface Antenna} 
\acrodef{ad}[AD]{Automatic Differentiation}
\acrodef{hris}[HRIS]{Hybrid Reconfigurable Intelligent Surface} 

\IEEEoverridecommandlockouts

\newcommand{\vecc}{{\operatorname{vec}}}



\title{Channel Estimation with Hybrid Reconfigurable Intelligent Metasurfaces}
\author{\IEEEauthorblockN{Haiyang Zhang,~\IEEEmembership{Member,~IEEE}, Nir Shlezinger,~\IEEEmembership{Member,~IEEE}, \\ George C. Alexandropoulos,~\IEEEmembership{Senior Member,~IEEE}, Avner Shultzman, \\Idban Alamzadeh, Mohammadreza F. Imani,~\IEEEmembership{Member,~IEEE}, \\and Yonina C. Eldar,~\IEEEmembership{Fellow,~IEEE}} 

\thanks{Parts of this work was presented at the IEEE SPAWC 2021 \cite{zhang2021channel}.
H. Zhang, A. Shultzman, and Y. C. Eldar are with the Faculty of Math and CS, Weizmann Institute of Science, Rehovot, Israel (e-mail: \{haiyang.zhang; avner.shultzman;  yonina.eldar\}@weizmann.ac.il). N. Shlezinger is with the School of ECE, Ben-Gurion University of the Negev, Beer-Sheva, Israel (e-mail: nirshl@bgu.ac.il). 
G.~C.~Alexandropoulos is with the Department of Informatics and Telecommunications, National and Kapodistrian University of Athens, 15784 Athens, Greece and also with the Technology Innovation Institute, 9639 Masdar City, Abu Dhabi, United Arab Emirates (e-mail:  alexandg@di.uoa.gr). 
I. Alamzadeh and M. F. Imani are with the School of ECEE, Arizona State University, Tempe, AZ, USA (email: \{amuham23; mohammadreza.imani\}@asu.edu).  This work was sponsored in part by the European Union’s H2020 research and innovation program under grant No. 101000967, in part by the Air Force Office of Scientific Research under grant No. FA9550-18-1-0208, in part by the Israel Science Foundation under grant No. 0100101, and in part by the EU H2020 RISE-6G project under grant number 101017011.}
	\vspace{-1.0cm}}
\vspace{-0.75cm}

\begin{document}
	
	\maketitle
\begin{abstract}
\acp{ris} are envisioned to play a key role in future wireless communications, enabling programmable radio propagation environments. They are usually considered as almost passive planar structures that operate as adjustable reflectors, giving rise to a multitude of implementation challenges, including the inherent difficulty in estimating the underlying wireless channels. In this paper, we focus on the recently conceived concept of \acp{hris}, which do not solely reflect the impinging waveform in a controllable fashion, but are also capable of sensing and processing an adjustable portion of it. We first present implementation details for this metasurface architecture and propose a convenient mathematical model for characterizing its dual operation. As an indicative application of \acp{hris} in wireless communications, we formulate the individual channel estimation problem for the uplink of a multi-user HRIS-empowered communication system. Considering first a noise-free setting, 
we theoretically quantify the advantage of \acp{hris} in notably reducing the amount of pilots needed for channel estimation, as compared to the case of purely reflective \acp{ris}. We then present closed-form expressions for the \ac{mse} performance in estimating the individual channels at the \acp{hris} and the base station for the noisy model.
Based on these derivations,  we propose an automatic differentiation-based first-order optimization approach to efficiently determine the \ac{hris} phase and power splitting configurations for minimizing the weighted sum-\ac{mse} performance. Our numerical evaluations demonstrate that \acp{hris} do not only enable the estimation of the individual channels in \ac{hris}-empowered communication systems, but also improve the ability to recover the cascaded channel, as compared to existing methods using passive and reflective \acp{ris}.
\end{abstract}
 \vspace{-0.3cm}



\acresetall

\vspace{-0.5cm}
\section{Introduction}
\acp{ris} are an emerging technology for the future $6$-th Generation (6G) of wireless communications, enabling dynamically programmable signal propagation over the wireless medium \cite{rise6g, RISE6G_COMMAG,huang2019reconfigurable, wu2019towards}. \acp{ris} are planar structures  comprised typically of multiple metamaterial elements, whose ElectroMagnetic (EM) properties can be externally controlled in a nearly passive manner, allowing them to realize various reflection and scattering profiles \cite{huang2020holographic}. By properly adjusting the reflection properties of the metamaterial elements, \acp{ris} can constructively strengthen/destructively weaken the desired/undesired signals at the target receiver(s). This ability of \acp{ris}  has been exploited in various promising communication systems, such as multi-user \acp{mimo} communications \cite{zheng2021double,huang2020achievable}, simultaneous wireless information and power transfer systems \cite{wu2020joint,pan2020intelligent}, and physical-layer security  \cite{pang2021irs}, for improving the respective performance. Achieving those performance gains with \acp{ris} often relies on accurate \ac{csi}. However, their passive nature implies that they can only act as adjustable reflectors, and thus, neither receive nor transmit their own data. This renders channel estimation a significant, but challenging task for \ac{ris}-based systems~\cite{swindlehurst2021channel}. 

In \ac{ris}-aided uplink communications, a signal transmitted from each \ac{ut} to the \ac{bs} undergoes at least two channels, namely, the \acp{ut}-\ac{ris} and \ac{ris}-\ac{bs} channels. With \acp{ris} being passive without any signal processing capability, the common approach to acquire \ac{csi} is to estimate only the entangled combined effect of the latter channels, i.e., the cascaded channel \cite{liu2019matrix,alwazani2020intelligent,chen2019channel} at the \ac{bs}. This can be achieved by having the \acp{ut} send known pilot symbols, which are reflected by the \ac{ris} such that the channel outputs at the \ac{bs} are used to estimate the overall channel. This approach has, however, two main drawbacks. First, since the number of reflective elements of \acp{ris} is usually very large, the cascaded channel is comprised of many unknown parameters that need to be estimated, which requires large pilot periods, and thus, significantly reduces the spectrum utilization efficiency. For example, for a system with $K$ UTs, $N$ \ac{ris} meta-atom elements, and $M$ \ac{bs} antennas, the cascaded channel consists of $KNM$ coefficients. To alleviate this drawback, several methods have been proposed to reduce the pilots, which impose a model on the overall channel having less coefficients, via, e.g.,  grouping the \ac{ris} elements  \cite{zheng2019intelligent}, exploiting the presence of a common channel \cite{wang2020channel}, and imposing a sparsity prior \cite{wang2020compressed}.  
The second drawback of passive RISs is that one
can only estimate the cascaded channel, instead of the individual ones,
which limits the plasticity for the transmission scheme design and restricts the network management flexibility \cite{hu2021two1}. As discussed in \cite{ye2020joint}, the individual UTs-RIS and RIS-BS channels are needed for some precoding designs. In addition, there exist certain scenarios where the RIS-BS channel may vary less rapidly than the UTs-RIS combined one. For those cases, only the UTs-RIS channel needs to estimated frequently, thus, it is desirable to be capable of recovering the UTs-RIS and RIS-BS channels individually~\cite{swindlehurst2021channel}.  

To overcome the aforementioned challenges with purely reflective \ac{ris}, it was recently proposed to equip \acp{ris} with minimal receive Radio-Frequency (RF) chains and antenna elements \cite{taha2021enabling, alexandropoulos2020hardware,jin2021channel,chen2021low}. In such architectures, some of the \ac{ris} elements are replaced with active receive antennas, which are connected via dedicated RF chains to a digital processor, and thus, have some signal processing capabilities such as reception and decoding. While such architectures enable the estimation of the individual \acp{ut}-\ac{ris} and \ac{ris}-\ac{bs} channels, they involve placing additional receivers along the \ac{ris}, possibly reducing its number of reflective elements. In addition, they are incapable of estimating the exact channel, since the signals observed at the \ac{ris} reflective elements are not measured, but only those acquired at the receive antenna elements. 


In parallel to the application of metasurfaces as passive reflective \acp{ris}, active metasurfaces have recently emerged as an appealing technology for realizing low-cost and low-power large-scale \ac{mimo} antennas \cite{shlezinger2020dynamic}. \acp{dma} pack large numbers of tunably radiative metamaterials on top of waveguides, resulting in \ac{mimo} transceivers with advanced analog processing capabilities \cite{shlezinger2019dynamic,wang2020dynamic,zhang2021beam,you2022EE_DMA,xu2022NF_DMA}. While the implementation of \acp{dma} differs from passive \acp{ris}, the similarity in the structure of the metamaterial elements between them indicates the feasibility of designing hybrid reflecting and sensing elements. This motivates studying the benefits from such a hybrid metasurface architecture, as an efficient means of facilitating \ac{ris}-empowered wireless communications, localization, and sensing. 


In our previous overview article \cite{alexandropoulos2021hybrid}, we proposed the \ac{hris} architecture, which is capable of simultaneously reflecting and receiving the incoming signal in an element-by-element controllable manner. \acp{hris} differ from both conventional passive \acp{ris}, which are usually metasurfaces operating in almost energy neutral tunable reflection, as well as from active \acp{dma} that operate similar to conventional transceiver antennas. In \acp{hris}, each metamaterial element enables simultaneous adjustable reflection and reception of the impinging signals. This controllable hybrid operation of \acp{hris} yields an architecture which generalizes \acp{ris} with interleaved reception and reflection meta-atoms, as in \cite{taha2021enabling,alexandropoulos2020hardware,jin2021channel,chen2021low}. 
More specifically, in \cite{alexandropoulos2021hybrid}, we overviewed the opportunities and challenges of the \ac{hris} concept, highlighting a hardware design for its implementation and presenting a full-wave-simulation-based proof-of-concept. However, no technical analysis relevant to the exploitation of the signal processing capability at the \ac{hris} side, and consequently of the availability of a portion of the received impinging signal, was provided. In this work, we fill this gap by considering \ac{hris}-assisted multi-user MIMO communication systems and investigating the individual channel estimation problem. We show that the reception signal processing capability of \acp{hris} allows the system to simultaneously estimate the individual channels, i.e., the \acp{ut}-\ac{hris} channel at the \ac{hris} side and the \ac{hris}-\ac{bs} at the \ac{bs}, via reusing the same transmitted pilot symbols. We first theoretically quantify the advantages of the \ac{hris} in terms of pilot reduction for the case without thermal noise. Then, considering the presence of this noise, we derive closed-form expressions for the \acp{mse} of the estimation of the individual channels at the \ac{hris} and \ac{bs}, respectively. Based on these derivations, we provide a gradient-based optimization approach to efficiently configure the \ac{hris} for minimizing the weighted sum-MSE, where the gradients of the complicated sum-MSE performance metric with respect to the HRIS tunable parameters are computed using \ac{ad}.

The main contributions of this paper are summarized as follows:
\begin{itemize}
\item \textit{Hybrid Reflecting and Sensing RIS Architecture: } 
We focus on the recently conceived concept of \acp{hris},
which enables metasurfaces to reflect the impinging  signal in an element-by-element controllable manner, while simultaneously sensing a portion of it. We specifically discuss the feasibility of hybrid meta-atom elements, present their high-level description, and provide a mathematical model for \ac{hris}-empowered wireless systems in a manner that is amenable to system design.
\item \textit{Estimation of the Individual Channels in \ac{hris}-Aided Systems: } 
We present an initial study on the potential gains of \acp{hris} in multi-user MIMO communication systems, by considering the individual channels estimation problem. 
We first characterize the number of pilots needed to estimate the channels for the case without noise,
and analytically demonstrate the gains of \acp{hris} compared to pure reflective \acp{ris}. We then consider typical noisy settings and derive the \ac{mse} for the estimation of the \acp{ut}-\ac{hris} channel at the \ac{hris} and the \ac{hris}-\ac{bs} channel at the \ac{bs}. 
\item \textit{\ac{ad}-based Gradient Optimization for \ac{hris} Configuration: } 
Since the \ac{mse} depends on the configuration of the \ac{hris}, we formulate a weighted sum-MSE minimization problem. The resulting problem is very challenging due to the complicated form of the objective and the large number of optimization variables. To deal with this, we propose a gradient-based solution to efficiently solve the problem. Inspired by the recent work \cite{shultzman2022nonlinear}, the gradients of the complicated sum-MSE objective function with respect to the HRIS parameters are computed analytically with \ac{ad}. The effectiveness of the proposed algorithm is verified numerically.
\item \textit{Numerical Evaluation:} Our simulation results showcase the inherent trade-off of HRISs concerning their ability to estimate the individual channels. Furthermore, it is demonstrated that, even when one aims to solely estimate the cascaded channel, \acp{hris} outperforms conventional (nearly) passive and reflective RISs \cite{wang2020channel} when the same pilot length is used.

\end{itemize}

The remainder of this paper is organized as follows. Section~\ref{sec:Model} presents the proposed model for generic \ac{hris} operation as well as the proposed \acp{hris}-assisted channel estimation approach. The individual channel estimation problem is investigated in Section~\ref{sec:Estimation}, which also includes the proposed gradient-based optimization for the \acp{hris} phase and power splitting parameters. Numerical evaluations are presented in Section~\ref{sec:Sims}, and Section~\ref{sec:Conclusions} provides  concluding remarks.

Throughout the paper, we use boldface lower-case and upper-case letters for vectors and matrices, respectively, while calligraphic letters are used for sets. The   vectorization operator,  transpose, conjugation, Hermitian transpose, trace, and  expectation are represented by ${\rm vec}(\cdot)$, $(\cdot)^T$, $(\cdot)^{\dag}$,  $(\cdot)^H$, ${\rm {Tr}}\left(\cdot\right)$, and $\E\{ \cdot \}$,  respectively. The notation ${\rm blkdiag}\left\{{\bf A}_1,{\bf A}_2,\ldots,{\bf A}_n\right\}$ denotes a block diagonal matrix with diagonal blocks given by ${\bf A}_1,{\bf A}_2,\ldots,{\bf A}_n$, and $[\mathbf{A}]_{i,j}$ denotes the $(i,j)$-th element of $\mathbf{A}$. Finally, $\mathbb{C}$ is the set of complex numbers.

\section{HRISs and System Modeling}
\label{sec:Model}
We first present a high-level description of the proposed hybrid metamaterial elements, followed by the considered system model for HRIS-empowered wireless  communications. We then detail a simple, yet convenient, model for the HRIS operation and present the proposed approach for the estimation of the individual channels.

\subsection{Hybrid Metasurfaces}
\label{subsec:HybridElements}
A rich body of literature has examined the fabrication of solely reflective \acp{ris} using metamaterials. A variety of implementations have been recently presented in \cite{huang2020holographic,Tsinghua_RIS_Tutorial}, ranging from RISs that change the wave propagation inside a multi-scattering environment for improving the received signal, to those which realize anomalous reflection, such that the reflected beam does not follow Snell's law and is directed towards desired directions. More recently, \acp{ris} which simultaneously refract and relfect their impinging signals were proposed to offer $360^\circ$ coverage \cite{liu2021star,mu2021simultaneously,xu2021star}.
In all those efforts, the RIS is not designed to sense the impinging signal. 

\begin{figure}
    \centering
    \includegraphics[width=0.43\columnwidth]{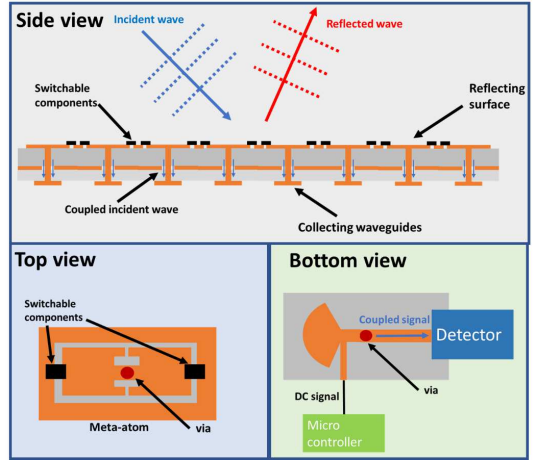}
    \caption{Illustration of a hybrid meta-atom which is capable of simultaneously reflecting a portion of its impinging wave in a reconfigurable manner, while feeding another portion of it to a receiving RF chain for baseband processing.}
    \label{fig:HybridAtom_v01}
\end{figure}

Metasurfaces can be designed to operate in a hybrid reflecting  and sensing manner. Such hybrid operation requires that each metasurface element is capable of simultaneously reflecting a portion of the impinging signal and receiving another portion of it in a controllable manner. As illustrated in Fig.~\ref{fig:HybridAtom_v01}, a simple mechanism for implementing such an operation is to couple each element to a waveguide. The signals coupled to the waveguides are then measured by receive RF chains and used to infer the necessary information about the channel. A detailed description of the practical implementation of such hybrid metamaterials can be found in \cite{alamzadeh2021reconfigurable}, where it was experimentally demonstrated that such surface configurations of hybrid metamaterials can be used to reflect in a reconfigurable manner, while using the sensed portion of the signal to locally recover its angle-of-arrival. In this paper, we are interested in examining the potential benefits of the HRIS paradigm in wireless communications. To this end, we present in the sequel a simple model capturing the simultaneous reflecting and sensing operation of HRISs, which is later on deployed to study HRIS-empowered wireless communications.

\subsection{HRIS Operation Modeling}
\label{subsec:HybridRIS}
\begin{figure}
		\centering			
		\includegraphics[width=0.7\columnwidth]{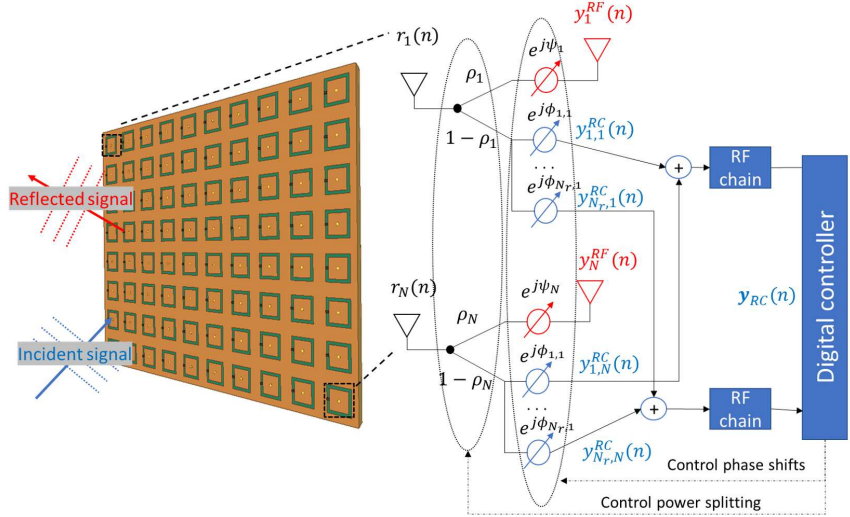}
		\vspace{-0.4cm}
		\caption{A simple model for the proposed \ac{hris} operation. The parameter $\rho_l$ models the portion of the impinging signal at the $l$-th meta-atom of the HRIS that gets tunably reflected, while $\psi_l$ and $\phi_{r,l}$ model the meta-atom's controllable phase shift and the joint effect of its response together with the analog phase shift before the $r$-th receive RF chain, respectively.} 
		\label{fig:Hybrid_RIS_model}
	\end{figure}

To model the dual reflection-reception operation of HRISs, we consider a hybrid metasurface comprised of $N$ meta-atom elements, which are connected to a digital controller via $N_r$ reception RF chains. Let $r_{l}(n)$ denote the radiation observed by the $l$-th HRIS element ($l=1,2,\ldots,N$) at the $n$-th time instance. A portion of this signal, dictated by the parameter $\rho_l \left( n \right) \in [0,1]$, is reflected with a controllable phase shift $\psi_l \left( n \right) \in [0,2\pi)$, and thus the reflected signal from the $l$-th element at the $n$-th time instant can be mathematically expressed as:
\begin{equation}\label{eq:reflection}
y_l^{\rm RF} (n)= \rho_l\left( n \right)e^{\jmath \psi_l \left( n \right)} r_l(n). 
\end{equation} 
The remainder of the observed signal is locally processed via analog combining and digital processing. The signal forwarded to the $r$-th RF chain via combining, with $r \in \{1,2,\ldots,N_r\}$, from the $l$-th element at the $n$-th time instant is consequently given by 
\begin{equation}\label{eq:power_split_t}
y_{r,l}^{\rm RC} (n) = (1 - \rho_l \left( n \right))e^{\jmath \phi_{r,l} \left( n \right)} r_l(n), 
\end{equation}
where $\phi_{r,l} \left( n \right)\in[0,2\pi)$ represents the adjustable phase that models the joint effect of the response of the $l$-th meta-atom and the subsequent analog phase shifting. The proposed HRIS operation model is illustrated in Fig.~\ref{fig:Hybrid_RIS_model}.

The operation of conventional passive and reflective RISs can be treated as a special case of the \ac{hris} architecture, by setting all $\rho_l \left( n \right)$ in \eqref{eq:reflection} equal to $1$. Compared with existing relay techniques, HRISs bring forth two major advantages. First, HRISs allow full-duplex operation (i.e., simultaneous reflection and reception) without inducing any self interference, which is unavoidable in full-duplex relaying systems. Second, HRISs require low power consumption since they do not need power amplifiers utilized by active transmit arrays; a typical receive RF consists of a low noise amplifier, a mixer which downconverts the signal from RF to baseband, and an analog-to-digital converter \cite{FD_MIMO}.
	
The resulting signal model at the HRIS can be expressed in vector form, as follows. By stacking the received signals $r_l(n), l=1,2,\ldots,N$ and the reflected signals $y_l^{\rm RF}(n), l=1,2,\ldots,N$ at the $N\times 1$ complex-valued vectors $\myVec{r}(n)$ and  $\myVec{y}_{\rm RF}(n)$, respectively, it follows from \eqref{eq:reflection} that:
\begin{equation} \label{eq:reflection_vector}
    \myVec{y}_{\rm RF}(n) = \boldsymbol{\Psi} \left(\boldsymbol{\rho}\left(n\right), \boldsymbol{\psi}\left(n\right)\right) \myVec{r}(n),
\end{equation}
with $\myMat{\Psi } \left( \boldsymbol \rho  \left( n \right), \boldsymbol \psi  \left( n \right)\right)  \triangleq {\rm diag} \left(\left[\rho_1  \left( n \right)\,e^{\jmath \psi_1  \left( n \right)},\rho_2  \left( n \right)\,e^{\jmath \psi_2  \left( n \right)},\dots,\rho_N  \left( n \right)\,e^{\jmath \psi_N  \left( n \right)}\right]\right)$.  
Similarly, by letting $\myVec{y}_{\rm RC}(n)\in \mathbb{C}^{N_r\times1}$ be the reception output vector at the HRIS, the following expression is deduced:
\begin{equation} \label{eq:output_compact}
    \myVec{y}_{\rm RC}(n) = {\boldsymbol \Phi}\left( \boldsymbol \rho  \left( n \right), \boldsymbol \phi  \left( n \right)\right) \myVec{r}(n),
\end{equation}
where the $N_r \times N$ matrix $\myMat{\Phi}\left( \boldsymbol \rho  \left( n \right), \boldsymbol \phi  \left( n \right)\right)$ represents the analog combining carried out at the \ac{hris} receiver. When the $l$-th meta-atom element is connected to the $r$-th RF chain, then $[\myMat{\Phi}\left( \boldsymbol \rho  \left( n \right), \boldsymbol \phi  \left( n \right)\right)]_{r,l} =  (1 - \rho_l \left( n \right))e^{\jmath \phi_{r,l} \left( n \right)}$, while when there is no such connection (e.g., for partially-connected analog combiners) it holds that
$[\myMat{\Phi}\left( \boldsymbol \rho  \left( n \right), \boldsymbol \phi  \left( n \right)\right)]_{r,l} = 0$. 

The reconfigurability of HRISs implies that the parameters $\boldsymbol \rho  \left( n \right)$ as well as the phase shifts $\boldsymbol \psi  \left( n \right)$ and $\boldsymbol \phi  \left( n \right)$ are externally controllable. It is noted that when an element is connected to multiple receive RF chains, then additional dedicated analog circuitry (e.g., conventional networks of phase shifters) is required to allow the signal to be forwarded with a different phase shift to each RF chain, at the possible cost of additional power consumption. Nonetheless, when each element feeds a single RF chain, then the model in Fig.~\ref{fig:Hybrid_RIS_model} can be realized without such circuitry by placing the elements on top of separated waveguides (see, e.g., \cite{shlezinger2020dynamic}).


\begin{figure}
    \centering
    \includegraphics[width=0.6\columnwidth]{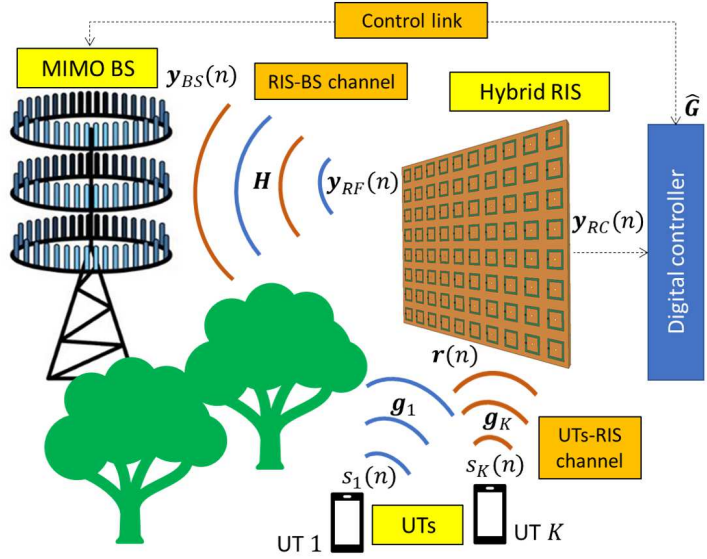}
    \caption{The considered \ac{hris}-empowered multi-user \ac{mimo} communication system operating in the uplink direction.}
    \label{fig:SystemModel1}
\end{figure} 
\subsection{HRIS-Assisted Channel Sounding}
\label{subsec:Problem}
In order to investigate the capabilities of HRISs in facilitating multi-user wireless communications, we henceforth study the problem of channel estimation in \ac{ris}-empowered systems, being one of the main challenges associated with conventional almost passive and reflective \acp{ris} \cite{hu2021two1,Tsinghua_RIS_Tutorial}. 
In particular, we consider an uplink multi-user \ac{mimo} system, where a \ac{bs} equipped with $M$ antenna elements serves $K$ single antenna \acp{ut} with the assistance of a \ac{hris}, as illustrated in Fig.~\ref{fig:SystemModel1}.
We assume that there is no direct link between the \ac{bs} and any of $K$ \acp{ut}, and thus communication is done only via the \ac{hris}. Let  $\myVec{H} \in \mathbb{C}^{M \times N}$ denote the channel gain matrix between the \ac{bs} and \ac{hris}, and ${\myVec{g}}_k \in \mathbb{C}^{N}$ be the channel gian vector between the $k$-th UT ($k=1,2,\ldots,K$) and \ac{hris}. We consider independent and identically distributed (i.i.d.) Rayleigh fading for all channels with $\myVec{H}$ and each ${\myVec{g}}_k$ having i.i.d. zero-mean Gaussian entries with variances $\beta$ and $\gamma_k$, respectively, denoting the path losses. In addition, for notation simplicity, we define  the matrix ${\myVec{G}}\triangleq\left[{\myVec{g}}_1,\cdots,{\myVec{g}}_K\right]$.
 
 \begin{figure}
    \centering
    \includegraphics[width=0.6\columnwidth]{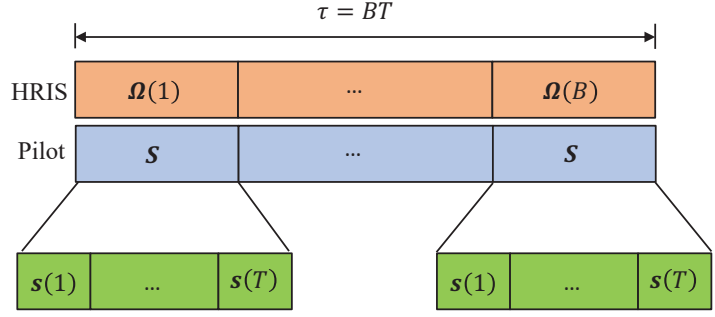}
    \caption{The frame structure for channel estimation using the proposed HRIS.}
    \label{fig:structure}
\end{figure}
 We consider a simple pilot-based channel training protocol, where the channel estimation time $\tau$ is divided into $B$ sub-frames, and each sub-frame consists of $T$ times slots such that $\tau = BT$, as depicted in Fig.~\ref{fig:structure}. The reconfigurable parameters of the HRIS remain constant during each sub-frame of $T$ time slots and vary from one sub-frame to another. Orthogonal pilot sequences $\left\{\myVec{s}_k \right\}_{k=1}^K$ are sent repeatedly over the $B$ sub-frames, where $\myVec{s}_k\triangleq\left[s_k\left(1\right),s_k\left(2\right),\ldots,s_k\left( T\right)\right] \in \mathbb{C}^{1 \times T}$ is the pilot sequence of the $k$-th \ac{ut} satisfying for for $1 \leq k_1,k_2 \leq K$: $\myVec{s}_{k_1}\myVec{s}_{k_2}^H = T$, if $k_1 = k_2$; and $\myVec{s}_{k_1}\myVec{s}_{k_2}^H = 0$, if $k_1 \neq k_2$. In Fig.~\ref{fig:structure}, $\myVec{s}(t) \triangleq \left[s_1\left( t\right),s_2\left( t\right),\ldots,s_K\left( t\right)\right]^T$ collects the pilot signals of the $K$ \acp{ut} at each $t$-th time slot for each sub-frame, and ${\boldsymbol \Omega}\left(b\right)\triangleq \left[{\boldsymbol \rho}(b), {\boldsymbol \phi}(b),{\boldsymbol \psi}(b)\right] $ includes all the optimization variables of the \ac{hris} at each $b$-th sub-frame. Consequently, the  signal received at the HRIS at each $t$-th time slot for each $b$-th sub-frame is given by:
 \begin{equation}\label{eq:received_all_pilot_output_new}
    \myVec{y}_{\rm RC}\left( b,t \right)={\boldsymbol \Phi}\left( {\boldsymbol \rho}(b), {\boldsymbol \phi}(b)\right) \myVec{G} \myVec{s}(t) + \myVec{z}_{\rm RC}\left( b,t \right), 
\end{equation}
where ${\boldsymbol \Phi}\left( {\boldsymbol \rho}(b), {\boldsymbol \phi}(b)\right)$ represents the reception matrix of the HRIS during the $b$-th sub-frame and
$\myVec{z}_{\rm RC}\left( b,t \right) \in\mathbb{C}^{N_r}$ is a zero-mean \ac{awgn} with entries having the variance $\sigma_{\rm RC}^2$. 
%
Similar to the derivation of \eqref{eq:received_all_pilot_output_new}, the signal received at the \ac{bs} at each $t$-th time slot for each $b$-th sub-frame can be expressed as:
\begin{equation}\label{eq:received_all_pilot_BS_new}
    \myVec{y}_{\rm BS}\left( b,t \right)= \myVec{H} { \boldsymbol  \Psi} \left( {\boldsymbol \rho}(b), {\boldsymbol \psi}(b)\right) \, \myVec{G} \myVec{s}(t) +  \myVec{z}_{\rm BS}\left( b,t \right),
\end{equation}
where $\myVec{z}_{\rm BS}\left( b,t \right) \in\mathbb{C}^{M }$ is a zero-mean \ac{awgn} having i.i.d. elements each with variance $\sigma_{\rm BS}^2$.

Let $\myVec{y}_{\rm RC}\left( b \right)\triangleq\left[\myVec{y}_{\rm RC}\left( b,1 \right),\myVec{y}_{\rm RC}\left( b,2 \right),\ldots,\myVec{y}_{\rm RC}\left( b,T \right)\right] \in \mathbb{C}^{N_r \times T}$ be the matrix collecting the received signals at the HRIS over $T$ time slots for each $b$-th block, i.e.:
\begin{equation}\label{eq:y_rc}
\myVec{y}_{\rm RC}\left( b \right) = {\boldsymbol \Phi}\left( {\boldsymbol \rho}(b), {\boldsymbol \phi}(b)\right) \myVec{G} \myVec{S} + \myVec{z}_{\rm RC}\left( b \right),
\end{equation}
with $\myVec{z}_{\rm RC}\left( b \right)\triangleq\left[\myVec{z}_{\rm RC}\left( b,1 \right),\myVec{z}_{\rm RC}\left( b,2 \right),\ldots,\myVec{z}_{\rm RC}\left( b,T \right)\right] \in \mathbb{C}^{N_r \times T} $ and $\myMat{S}\triangleq\left[\myVec{s}(1),\myVec{s}(2),\ldots,\myVec{s}(T) \right]$, where it holds that $\myMat{S}\myMat{S}^H = T{\bf I}_K$. We, then, define $\myVec{y}_{\rm RC}$ as the $N_r\, B\times T$ matrix generated by stacking the rows of the $B$ matrices $\myVec{y}_{\rm RC}\left( 1 \right),\myVec{y}_{\rm RC}\left( 2\right),\ldots, \myVec{y}_{\rm RC}\left( B\right)$.
 From \eqref{eq:y_rc}, $\myVec{y}_{\rm RC}$ can be written as a linear function of the \acp{ut}-\ac{hris} channel $\myVec{G}$ as 
\begin{equation}
    \label{eqn:LinearRC}
    \myVec{y}_{\rm RC} = \myMat{A}_{\rm RC}\left( \left\{\myVec{\rho}(b), \myVec{\phi}(b)\right\}\right) {\myVec{G}} \myVec{S} + \myVec{z}_{\rm RC},
\end{equation}
where $\myVec{z}_{\rm RC} \in \mathbb{C}^{N_r\, B\times T}$ results from the row stacking of the matrices $\myVec{z}_{\rm RC}(1),\myVec{z}_{\rm RC}(2),\ldots, \myVec{z}_{\rm RC}(B)$, while the matrix $\myMat{A}_{\rm RC} \in \mathbb{C}^{N_r\, B\times N}$ is defined as:
\begin{equation}\label{eq:Arc}
   \myMat{A}_{\rm RC}\left( \left\{\myVec{\rho}(b), \myVec{\phi}(b)\right\}\right) \triangleq \left[\myMat{\Phi}\left( \myVec{\rho}(1), \myVec{\phi}(1)\right)^T,\cdots,\myMat{\Phi}\left( \myVec{\rho}(B), \myVec{\phi}(B)\right)^T\right]^T.
\end{equation}
Similarly, by letting $\myVec{y}_{\rm BS}\left( b \right)\triangleq\left[\myVec{y}_{\rm BS}\left( b,1 \right),\myVec{y}_{\rm BS}\left( b,2 \right),\ldots,\myVec{y}_{\rm BS}\left( b,T \right)\right] \in \mathbb{C}^{M \times T}$ be the matrix including  the received signals at BS during the $T$ time slots for each $b$-th sub-frame, we have:
\begin{equation}\label{eq:y_BS}
\myVec{y}_{\rm BS}\left( b \right) = \myVec{H} {\boldsymbol \Psi}\left( {\boldsymbol \rho}(b), {\boldsymbol \psi}(b)\right) \myVec{G} \myVec{S} + \myVec{z}_{\rm BS}\left( b \right),
\end{equation}
where $\myVec{z}_{\rm BS}\left( b \right)\triangleq\left[\myVec{z}_{\rm BS}\left( b,1 \right),\myVec{z}_{\rm BS}\left( b,2 \right),\ldots,\myVec{z}_{\rm BS}\left( b,T \right)\right] \in \mathbb{C}^{M \times T} $.

As in conventional \ac{ris}-empowered communication systems, e.g., \cite{huang2019reconfigurable,wu2019intelligent}, we assume that the \ac{bs} maintains a high-throughput direct link with the \ac{hris}. For passive \acp{ris}, this link is used for controlling the \ac{ris} reflection pattern. In \acp{hris}, which have reception, thus measurement collection, capabilities, this link is also used for conveying valuable information from the \ac{hris} to the \ac{bs}. Therefore, inspired by the recent discussions for autonomous RISs with basic computing and storage capabilities \cite{self_configuring_RIS,DRL_automonous_RIS}, we focus on channel estimation carried out at both the \ac{hris} side as well as the \ac{bs}. Our goal is to characterize the achievable \ac{mse} in recovering the \acp{ut}-\ac{hris} channel $\myVec{G}$ from \eqref{eqn:LinearRC}, along with the \ac{mse} in estimating $\myVec{H}$ at the \ac{bs} from \eqref{eq:y_BS} and from the estimate of $\myVec{G}$, denoted $\hat{\myVec{G}}$, provided by the \ac{hris}. The pilot matrix $\myVec{S}$ in \eqref{eqn:LinearRC} and \eqref{eq:y_BS} is assumed to be known at both the \ac{hris} and \ac{bs}. It ia also noted that different \ac{hris} configurations $\left\{\myVec{\rho}(b), \myVec{\phi}(b), \myVec{\psi}(b)\right\}$ result in different pilot signal strengths in \eqref{eqn:LinearRC} and \eqref{eq:y_BS}. Therefore, we also aim at configuring the \ac{hris} controllable parameters in order to facilitate channel estimation based on the characterized \ac{mse}.


\section{Estimation of the Individual Channels}
\label{sec:Estimation}
In this section, we quantify the HRIS potential in facilitating estimation of all individual channels in the uplink of HRIS-empowered multi-user MIMO communication systems.  In the considered model, the individual \acp{ut}-\ac{hris}  channel is estimated at the \ac{hris} side based on \eqref{eqn:LinearRC}, while the individual \ac{hris}-\ac{bs} channel  is estimated at the \ac{bs} from \eqref{eq:y_BS} using the estimation $\hat{\myVec{G}}$, which is provided by the \ac{hris}.
In this section, we first study the number of pilots needed to estimate the channels for a noise-free setting in Subsection~\ref{subsec:Noiseless}, where $\myMat{G}$ and $\myMat{H}$ can be identified with no errors. Then, we express the achievable channel estimation \ac{mse} for noisy reception in Subsection~\ref{subsec:Noisy}, which we then use to optimize the \ac{hris} in Subsection~\ref{sec:graph-based}. A discussion on the proposed channel estimation approach is provided in Subsection~\ref{subsec:Discussion}.

\subsection{Channel Estimation for Noise-Free Channels}
\label{subsec:Noiseless}
We begin by considering communications carried out in the case of without noise, where the noise terms in \eqref{eq:received_all_pilot_output_new} and \eqref{eq:received_all_pilot_BS_new} are set to be zero, i.e., $\sigma_{\rm RC}^2=\sigma_{\rm BS}^2=0$. In such scenarios, one should be able to fully recover both $\myVec{H}$ and $\myVec{G}$ from the observed signals $\myVec{y}_{\rm RC}(n)$ and $\myVec{y}_{\rm BS}(n)$. The number of pilots required to achieve  accurate recovery is stated in the following proposition.
%
\begin{proposition}
\label{pro:Noiseless}
In the case of without noise, $\myVec{H}$ and $\myVec{G}$ can be accurately recovered when the number of pilots $\tau$ satisfies the inequality:
\begin{equation}
    \label{eqn:Noiseless}
    \tau \geq N \cdot \max\left\{1,KN_r^{-1}\right\}.
\end{equation}
\end{proposition}
\ifFullVersion
\begin{IEEEproof}
The proof is provided in Appendix \ref{app:Proof1}.  
\end{IEEEproof}
\fi

Proposition \ref{pro:Noiseless} demonstrates the intuitive benefit of \acp{hris} in facilitating individual channel estimation with reduced number of pilots, as compared to existing techniques for estimating the cascaded \acp{ut}-\ac{ris}-\ac{bs} channels (e.g., \cite{wang2020channel}). For instance, for a multi-user \ac{mimo} system with $M=16$ \ac{bs} antennas, $N_r=8$ \ac{hris} RF chains, $K=8$ \acp{ut}, and $N=64$ \ac{hris} elements, the adoption of an \ac{hris} allows recovering $\myVec{H}$ and $\myVec{G}$ separately using $\tau=64$ pilots. By contrast, the method proposed in \cite{wang2020channel} requires transmitting over $90$ pilots to identify the cascaded channel coefficients $[\myVec{H}]_{m,l} [\myVec{G}]_{l,k}$  for $l,k$ and $m=1,2,\ldots,M$. This reduction in pilot signals is directly translated into improved spectral efficiency, as less pilots are to be transmitted in each coherence duration.

\subsection{Channel Estimation for Noisy Channels}
\label{subsec:Noisy}
The characterization of the number of required pilots in Proposition \ref{pro:Noiseless} provides an initial understanding of the \ac{hris}'s capability in providing efficient channel estimation. However, as Proposition \ref{pro:Noiseless} considers an effectively noise-free setup, it is invariant of the fact that \acp{hris} split the power of their received signal $\myVec{r}(n)$ between the reflected and received components. In the presence of noise, this division of the signal power may result in \ac{snr} degradation. Therefore, we next study  channel estimation using \acp{hris} in the presence of noise, quantifying the achievable \ac{mse} in estimating the individual \acp{ut}-\ac{hris} and \ac{hris}-\ac{bs} channels for a fixed \ac{hris} configuration $\left\{\myVec{\rho}(b), \myVec{\phi}(b), \myVec{\psi}(b)\right\}$. 
For notation brevity, in the following, we define the set of \ac{hris} parameters affecting its reception as $\myMat{\Phi} \triangleq \left\{\myVec{\rho}(b), \myVec{\phi}(b)\right\}$, the \ac{hris} parameters affecting the reception at the \ac{bs} as $\myMat{\Psi}(b) \triangleq \myMat{\Psi}(\myVec{\rho}(b),\myVec{\psi}(b))$, and the overall \ac{hris} parameters as $\myMat{\Omega} \triangleq \left\{ \myVec{\rho}(b), \myVec{\phi}(b), \myVec{\psi}(b)\right\}$. We also make the assumption that the noise powers at the \ac{hris} and \ac{bs} are of the same level, i.e., $\sigma_{\rm BS}^2 = \sigma_{\rm RC}^2=\sigma^2$.
We begin by characterizing the achievable \ac{mse} performance in recovering $\myVec{G}$ at the \ac{hris} using its locally collected measurements for a given \ac{hris} parameterization, denoted by $\mathcal{ E}_{\myVec {G}} \left( \myMat{\Phi} \right)$.
\begin{theorem}\label{Theorem:1}
The \acp{ut}-\ac{hris} channel $\myVec{G}$ can be recovered with the following \ac{mse} performance:
\begin{equation*}
\mathcal{ E}_{\myVec{G}} \left( \myMat{\Phi} \right)  = {\rm Tr} \left\{\left(\myVec{R}_{\rm G}^{-1}\, +T\frac{ \Gamma}{K}\myMat{A}_{\rm RC}\left( \myMat{\Phi} \right)^H\, \myMat{A}_{\rm RC}\left( \myMat{\Phi} \right)\right)^{-1}\right\}
\end{equation*}
where $\myVec{R}_{\rm G} \triangleq \left(\sum_{k=1}^K \gamma_k\right) \myVec{I}_N$;  $\Gamma\triangleq\frac{P_t}{\sigma^2}$ represents the transmit SNR with $P_t$ denoting each UT's power used for transmitted the pilot symbols.
\end{theorem}
\ifFullVersion
\begin{IEEEproof}
The proof is given in Appendix \ref{app:Proof2}.
\end{IEEEproof}
\fi

Theorem~\ref{Theorem:1} allows to compute the achievable \ac{mse} in estimating $\myVec{G}$ at the \ac{hris} side for a given configuration of its reception phase profile, determined by $\myMat{\Phi} $, i.e., by $ \left\{\myVec{\rho}(b)\right\}$ and $\left\{\myVec{\phi}(b)\right\}$.

\begin{lemma}\label{lemma:1}
Let $\myVec{\hat G}$ and $\myVec{\tilde G} = \myVec{G} - \myVec{\hat G}$ denote the estimation of $\myVec{G}$ and its corresponding estimation error, respectively. From Appendix \ref{app:Proof2}, the distributions of $\myVec{\hat G}$ and $\myVec{\tilde G}$ are respectively given by
\begin{equation*}
\myVec{\hat G}    \sim \mathcal{C} \mathcal{N} \left(0,{\boldsymbol \Sigma} \left(\myMat{\Phi}\right) \right), \qquad
\myVec{\tilde G}    \sim \mathcal{C} \mathcal{N} \left(0,\myVec{R}_{\bf \tilde G} \left(\myMat{\Phi}\right)\right),
\end{equation*}
where $\myVec{R}_{\bf \tilde G} \left(\myMat{\Phi}\right) $ and ${\boldsymbol \Sigma} \left(\myMat{\Phi}\right) $ are defined as
$\myVec{R}_{\bf \tilde G} \left(\myMat{\Phi}\right)  \triangleq  
\left(\myVec{R}_{\rm G}^{-1}\, + T \Gamma K^{-1}\,\myMat{A}_{\rm RC}\left( \myMat{\Phi}\right)^H\, \myMat{A}_{\rm RC}\left( \myMat{\Phi}\right)\right)^{-1}$ and ${\boldsymbol \Sigma} \triangleq \myVec{R}_{\rm G}\,  \myMat{A}_{\rm RC}\left( \myMat{\Phi}\right)^H\,
 \Bigg( \myMat{A}_{\rm RC}\left( \myMat{\Phi}\right)\, \myVec{R}_{\rm G}\, \myMat{A}_{\rm RC}\left(\myMat{\Phi}\right)^H + K\,  (T \Gamma)^{-1}\,{\myVec{I}_{N_rB}} \Bigg)^{-1} \myMat{A}_{\rm RC}\left( \myMat{\Phi}\right) \myVec{R}_{\rm G}^H$.
 \end{lemma}


Lemma \ref{lemma:1} provides the statistical results for the estimation of $\myVec{ G}$ with respect to the reconfigurable parameters $\myMat{\Phi} $; more specifically, to $ \left\{\myVec{\rho}(b)\right\}$ and $\left\{\myVec{\phi}(b)\right\}$. This estimation can be used from the BS to estimate the channel matrix $\myMat{H}$. In particular, letting the \ac{hris} convey the estimation of $\myMat{G}$ to the \ac{bs} (via their control link) allows achieving the \ac{mse} in recovering $\myVec{H}$, as described by means of the following theorem. Recall that the BS observes the reflected portion of the signal at the output of the \ac{hris}-\ac{bs} channel.

\begin{theorem}\label{Theorem:2}
The \ac{hris}-\ac{bs} channel $\myVec{H}$ can be recovered with the following \ac{mse} performance: 
%
\begin{align*}  
&\mathcal{E}_{\myVec{H}}\left(\myMat{\Omega} \right) = {\rm Tr} \left(\left(\frac{1}{\beta} \myMat{I}_{M N} + \left(K \sum_{j=1}^B \sum_{i=1}^B\,{\rm Tr}\left(\left[\myVec{D}(\myMat{\Omega})^{-T}\right]_{i,j}\right)\, \myMat{\Psi}(i)\,{\boldsymbol \Sigma} \left(\myMat{\Phi}\right) \myMat{\Psi}(j)^H \right)^T
\otimes \myVec{I}_M\right)^{-1}\right),
\end{align*}
where $\myVec{D}(\myMat{\Omega})$ is a ${BK \times BK}$ matrix which can be partitioned into $B \times B$ blocks with each block being a $K \times K$ submatrix. The $i$-th row and $j$-th column block of $\myVec{D}(\myMat{\Omega})$ is defined as:
\begin{equation*}
\begin{split}
&\left[\myVec{D}\left(\myMat{\Omega} \right)\right]_{i,j} = \left\{\begin{array}{ll}
\frac{\beta}{K} {\rm Tr}\left(\myMat{\Psi}(j)^H\myMat{\Psi}(i)\myVec{R}_{\bf \tilde G} \left(\myMat{\Phi}\right)\right)\myVec{I}_K + (T \Gamma)^{-1} \myVec{I}_{ K}, & {\rm if}~ i = j \\
\frac{\beta}{K} {\rm Tr}\left(\myMat{\Psi}(j)^H\myMat{\Psi}(i)\myVec{R}_{\bf \tilde G} \left(\myMat{\Phi}\right)\right)\myVec{I}_K, & {\rm if}~ i \neq j
\end{array}\right..
\end{split}
\end{equation*}
\end{theorem}

\ifFullVersion
\begin{IEEEproof}
The proof is provided in Appendix~\ref{app:Proof3}.
\end{IEEEproof}
\fi

Theorems~\ref{Theorem:1} and \ref{Theorem:2}
allow us to evaluate the achievable \ac{mse} for the recovery of the individual channels $\myVec{G}$ and $\myVec{H}$. The fact that these \acp{mse} are given as functions of the \ac{hris} parameters $\myMat{\Omega}$ enables us to numerically optimize the HRIS configuration for the estimation of the individual channels. 
In Section~\ref{sec:Sims}, our numerical evaluation of the \ac{mse} performance reveals the fundamental trade-off between the ability to recover $\myVec{G}$ and $\myVec{H}$, which is dictated mostly by the parameter $\myVec{\rho}$ determining the portion of the impinging signal being reflected; the remaining portion is sensed and used for channel estimation at the HRIS side. 


\subsection{\ac{hris} Configuration Optimization}
\label{sec:graph-based}
The formulation of the channel estimation \ac{mse} given previously for a fixed \ac{hris} configuration motivates the optimization of its parameters. We henceforth seek to optimize the HRIS parameters $\myMat{\Omega} = \left\{\boldsymbol \rho \left(b \right),  \boldsymbol \psi \left(b \right), \boldsymbol \phi \left(b \right)\right\}$ so as to minimize the weighted-sum MSE. Mathematically, the optimization problem under investigation is formulated as follows:
 %
\begin{equation} \label{eq:original_problem_new}
\begin{split}
&\min_{ \left\{\boldsymbol \rho \left(b \right),  \boldsymbol \psi \left(b \right), \boldsymbol \phi \left(b \right)\right\}}~~  \mathcal{E}_{\myVec{H}}\left( \left\{\boldsymbol \rho \left(b \right),  \boldsymbol \psi \left(b \right), \boldsymbol \phi \left(b \right)\right\} \right) +\mathcal{ E}_{\myVec{G}} \left( \left\{\boldsymbol \rho(b), \boldsymbol \phi(b) \right\}\right)\\
&~~~~~~s.t.~~~~~~~~~~[\myMat{\rho}(b)]_p \in [0, 1], ~[\myMat{\psi}(b)]_p \in [0,2\pi],~[\myMat{\phi}(b)]_q \in [0,2\pi],~\\
&~~~~~~~~~~~~~~~~~~~~b=1,2,\ldots,B,~ p=1,2,\ldots,N, ~ q=1,2,\ldots,N\times N_r.
\end{split}
\end{equation} 
Problem \eqref{eq:original_problem_new} is non-convex and challenging to solve, even when using numerical approaches based on Bayesian optimization, as previously proposed for \ac{ris} configuration in complex settings \cite{wang2021jointly}. This is because the dimension of the optimization variables is extremely high when the number of antennas at the BS and the number of meta-atom elements at the \ac{hris} are large. 
%
%
%
%
%
We thus propose to tackle the \ac{hris} reconfiguration problem in  \eqref{eq:original_problem_new} via a gradient-based optimization approach. 

The gradients of the sum-MSE objective function with respect to the HRIS parameters can be computed analytically with \ac{ad} \cite{Griewank1989automaticDifferentiation}, which is extensively used for machine learning applications. To compute the gradients of a differentiable function automatically, \ac{ad} expresses the function as a computational graph and applies the backpropagation algorithm to retrieve the gradients. To justify the use of \ac{ad}, we recall that the sum-MSE objective function is composed of basic differentiable operations, such as the matrix trace operation. 
Moreover, the composition of differentiable functions results in a differentiable function, and thus the sum-MSE objective function is differentiable, allowing us to apply \ac{ad}.

For notation brevity, in the following, we define $f\left(\mathbf{x}\right)$ as the objective function of \eqref{eq:original_problem_new} with $\mathbf{x}\triangleq\left[{\boldsymbol \rho }\left(1 \right),\ldots,\boldsymbol \rho \left(B \right),{{\boldsymbol \psi} }\left(1 \right),\ldots,{\boldsymbol \psi} \left(B \right),{{\boldsymbol \phi} }\left(1 \right),\ldots,{\boldsymbol \phi} \left(B \right)\right]$ including the elements of the set $\myMat{\Omega}$ of the \ac{hris} parameters in a vector form. To deal with the inequality constraints on the HRIS free variables, we add a barrier regularization term to the objective function of \eqref{eq:original_problem_new}, resulting in the following optimization problem:
\begin{equation} \label{eq:unconstrained_optimization_with_barrier}
    \min_{\mathbf{x}\in  \mathcal{C}}~~~ \mathcal{L}(\mathbf{x}) \triangleq f(\mathbf{x}) + \lambda \mathcal{B}_{\mathcal{C}}(\mathbf{x}).
\end{equation}
where $\lambda \in \mathbb{R}$ is a regularization hyperparameter, $ \mathcal{C}$
denotes the feasible set of $\mathbf{x}$, given by equation $ \mathcal{C} \triangleq \left\{ \mathbf{x} \in \mathbb{R}^{BN\left( 2  +  N_r\right)} : 0 \leq \mathbf{x}_i \leq 1, \forall i \in \left[1, BN\right]~{\rm and}~ 0 \leq \mathbf{x}_i \leq 2\pi, \forall i \in \left[BN+1, BN\left( 2  +  N_r\right)\right] \right\}$, and $\mathcal{B}_{\mathcal{C}}(\mathbf{x})$ represents the barrier function, which adds a high penalty to the points approaching the feasible region's boundaries.
Formally, a barrier function $\mathcal{B}_{\mathcal{C}}(\mathbf{x})$ is any function that satisfies: \textit{i}) $ \mathcal{B}_{\mathcal{C}}(\mathbf{x}) \geq 0$ $\forall \mathbf{x} \in \mathcal{C}$; and \textit{ii}) $\mathcal{B}_{\mathcal{C}}(\mathbf{x}) \rightarrow \infty$ $\forall \mathbf{x} \rightarrow \partial \mathcal{C}$, where $\partial \mathcal{C}$ denotes the boundaries of the feasible region. In the problem formulation in \eqref{eq:unconstrained_optimization_with_barrier}, we adopt the barrier function:
\begin{equation} \label{eq:barrier_sum_MSE}
    \mathcal{B}_{\mathcal{C}}(\mathbf{x}) =  \sum_{i=1}^{BN} \left(\frac{1}{\mathbf{x}_i} + \frac{1}{1 - \mathbf{x}_i}\right) + \sum_{i=BN+1}^{BN\left( 2  +  N_r\right)} \left(\frac{1}{\mathbf{x}_i} + \frac{1}{2\pi - \mathbf{x}_i}\right),
\end{equation}
which is differentiable, returns non-negative values in the feasible set, and is not bounded as the variables approach the feasible set's boundaries. To solve the resulting minimization problem, we apply a gradient-based iterative approach, where at each iteration we compute the derivative with an AD tool (we used PyTorch's autograd engine \cite{NEURIPS2019_9015}) and update the parameters with a first-order optimizer (e.g., gradient descent and its variants, such as Adam \cite{Kingma2014Adam}).
The solution using a conventional gradient descent algorithm with AD is summarized in Algorithm~\ref{alg:computational_graph_optimization}.
 \begin{algorithm}[!t]
 \caption{\ac{hris} Configuration for Weighted-Sum MSE Minimization}
 \label{alg:computational_graph_optimization}
 \begin{algorithmic}[1]
  \renewcommand{\algorithmicrequire}{\textbf{Initialize:}} \REQUIRE  $\mathbf{x}^{(0)}$, step-size $\eta$, and $t \leftarrow 0$.
  \WHILE{stopping criteria is not satisfied}
  \STATE Update the objective value $\mathcal{L}(\mathbf{x})$ in  \eqref{eq:unconstrained_optimization_with_barrier}.\\
  \STATE Compute the gradients using the AD-based backpropagation algorithm $\nabla_{\mathbf{x}} \mathcal{L}(\mathbf{x}^{(t)})$.\\
  \STATE Update the parameters vector via $\mathbf{x}^{(t+1)}\leftarrow\mathbf{x}^{(t)}-\eta\nabla_{\mathbf{x}} \mathcal{L}(\mathbf{x}^{(t)})$. \\
    \STATE Update $t\leftarrow t+1$.
  \ENDWHILE
 \renewcommand{\algorithmicrequire}{\textbf{Output:}} \REQUIRE  $\left\{\boldsymbol \rho \left(b \right),  \boldsymbol \psi \left(b \right), \boldsymbol \phi \left(b \right)\right\}\leftarrow  \mathbf{x}^{(t)}$.
 \end{algorithmic} 
 \end{algorithm}
\subsection{Discussion}
\label{subsec:Discussion}
The fact that \acp{hris} require less pilots naturally follows from their ability to provide additional $N_r$ reception ports, while simultaneously acting as a dynamically configurable reflector.  It is noted that our results in the previous subsections are obtained assuming that the \acp{ut}-\ac{hris} channel $\myVec{G}$ is estimated at the \ac{hris}, and its estimate is then forwarded to the \ac{bs}. 
Furthermore, exploiting the \ac{hris} as an additional non-co-located receive port can also facilitate data transmission once the channels are estimated. Though, in this case, one would also have to account for possible rate limitations on the \ac{hris}-\ac{bs} link. We leave the study of these additional usages of \acp{hris} for future research. 


The study of \acp{hris}, combined with the numerical evaluations in Section~\ref{sec:Sims} that follows, only reveal a portion of the potential of  \acp{hris} in facilitating wireless communication over programmable environments.  To further understand the contribution of \acp{hris}, one should also study their impact on data transmission, as well as consider the presence of an additional direct channel between the \acp{ut} and the \ac{bs}. Furthermore, the simplified model used in this work is based on the hybrid metamaterial model 
presented in \cite{alexandropoulos2021hybrid}. To this end, additional experimental studies of this model are required to formulate a more accurate physically-compliant model for the behavior of \acp{hris}; see \cite{PhysFad} and references therein for recent modeling research. These extensions are also left for future work.


Our proposed hybrid \ac{ris} structure is different from another emerging hybrid \ac{ris} concept, called simultaneously transmitting and reflecting (STAR)-\ac{ris} \cite{xu2021star,mu2021simultaneously,liu2021star}.
In STAR-RIS, each antenna element splits its incident signal power into two parts, i.e., one part of the signal is reflected in the same space as the incident signal, and the other part of the signal is transmitted to the opposite space of the incident signal. Hence, STAR-RIS is capable of simultaneous reflection and refraction (passing the signal through the surface to its other side), without having any receiving or decoding capability.


\section{Numerical Results}
\label{sec:Sims}

In this section, we numerically evaluate the performance of the proposed channel estimation approach for the uplink of \ac{hris}-empowered multi-user MIMO communication systems. Specifically, we present simulation parameters in Section~\ref{subsec:simulation} and then provide our numerical results in Section~\ref{subsec:simu}.

\subsection{Simulation Setup}
\label{subsec:simulation}
 In our simulations, the pathlosses of the individual channels ${\bf H}$ and ${\bf g}_k$ are modeled as $\beta = \lambda_0 \left(\frac{d_H}{d_0}\right)^{-\alpha_h}$ and $\gamma_k = \lambda_0\left(\frac{d_k}{d_0}\right)^{-\alpha_g}$, respectively, where $\lambda_0 = -20$ dB denotes a constant pathloss at the reference distance $d_0 = 1$ m, while $d_H$ and $d_k$ are the distances from the \ac{hris} to the \ac{bs} and $k$th \ac{ut}, respectively. The pathloss exponents were set as $\alpha_h = 2.2$  and $\alpha_g = 2.1$. The above wireless channel parameters were also adopted in  \cite{wang2020channel}. We consider a 2D Cartesian coordinate system in which the BS and the \ac{hris} are respectively located at points (0, 0) and (0, 50 m), while the $K$ users were randomly generated in an area centered at (30 m, 50 m) with a radius of $10$ m.  In addition, we have set the numbers of antennas at the \ac{bs} and the number of meta-atom elements at the \ac{hris} as $M=16$ and $N=64$, respectively, and the number of \acp{ut} as $K=8$, unless otherwise stated.

\begin{figure}
    \centering
    \includegraphics[width=0.7\columnwidth]{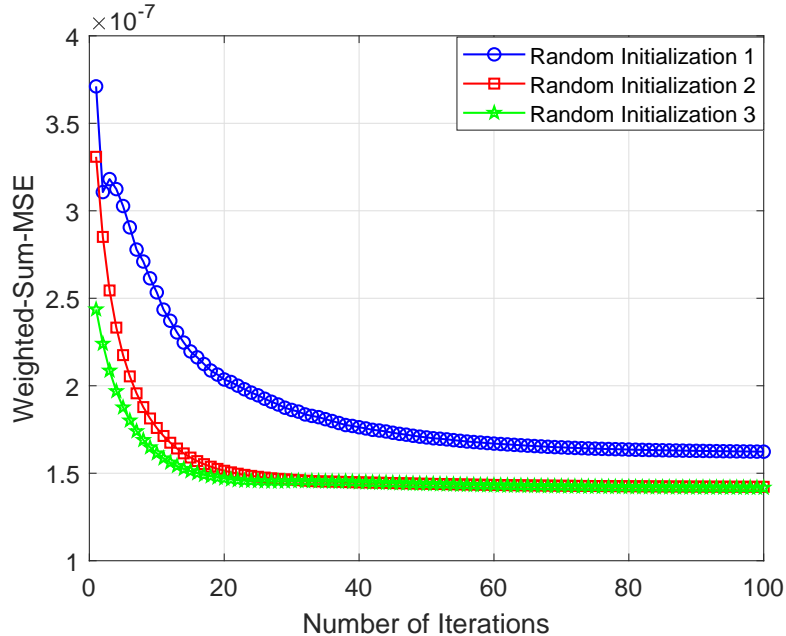}
    \caption{Convergence behavior of the weighted-sum MSE for the proposed channel estimation algorithm with different random initialization using $N_r=8$ receive RF chains at the HRIS, transmit SNR of $\Gamma = 100$~dB, and $\tau = 104$ pilots symbols.
    }
    \label{fig:convergence}
\end{figure}


\subsection{Simulation Results}
\label{subsec:simu}
We first demonstrate the convergence behavior of our proposed \ac{ad}-based Algorithm~\ref{alg:computational_graph_optimization} for solving the considered channel estimation problem. Specifically, Fig.~\ref{fig:convergence} depicts the convergence of the achievable weighted sum-MSE for different random initialization of the optimization variables, when the number of receive RF chains at the \ac{hris} is set to $N_r=8$, the transmit \ac{snr} $\Gamma = 100$~dB, and the pilot length is $\tau = 104$. It can be observed that for each randomly generated initialization, the proposed algorithm converges to a fixed value within $100$ iterations, and typically with much fewer iterations, verifying its relatively fast convergence. Therefore, we have set the maximum number of iterations to be $100$ for Algorithm~\ref{alg:computational_graph_optimization} in the following simulation experiments.
\begin{figure} 
  \centering 
    \subfigure[Normalized \ac{mse} of estimating the UTs-HRIS channel ${\bf G}$.]{ 
    \label{fig:subfig:G} 
    \includegraphics[width=3.1in]{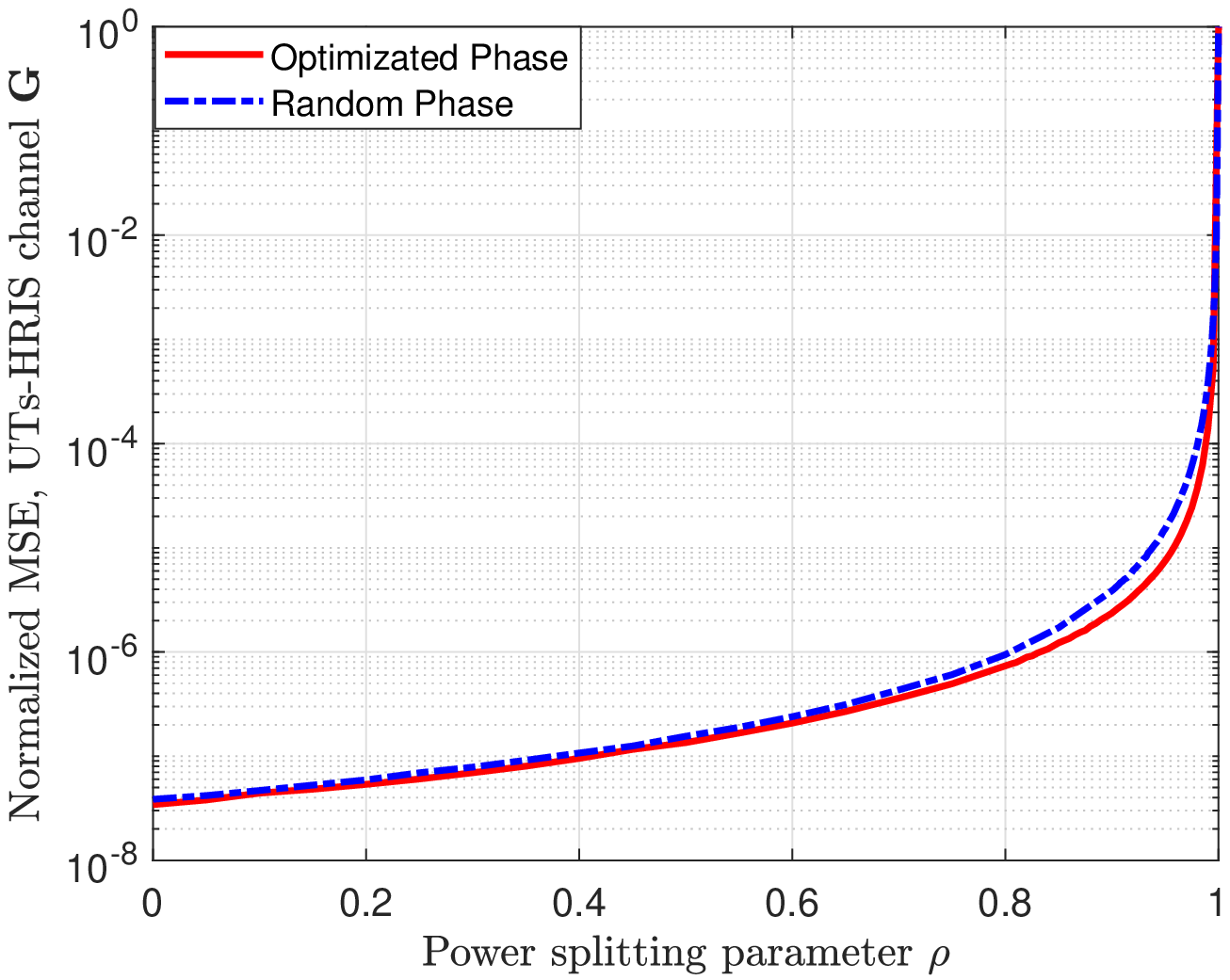}
    } 
  \subfigure[Normalized \ac{mse} of estimating the HRIS-\ac{bs} channel ${\bf H}$.]{ 
    \label{fig:subfig:H}
    \includegraphics[width=3.1in]{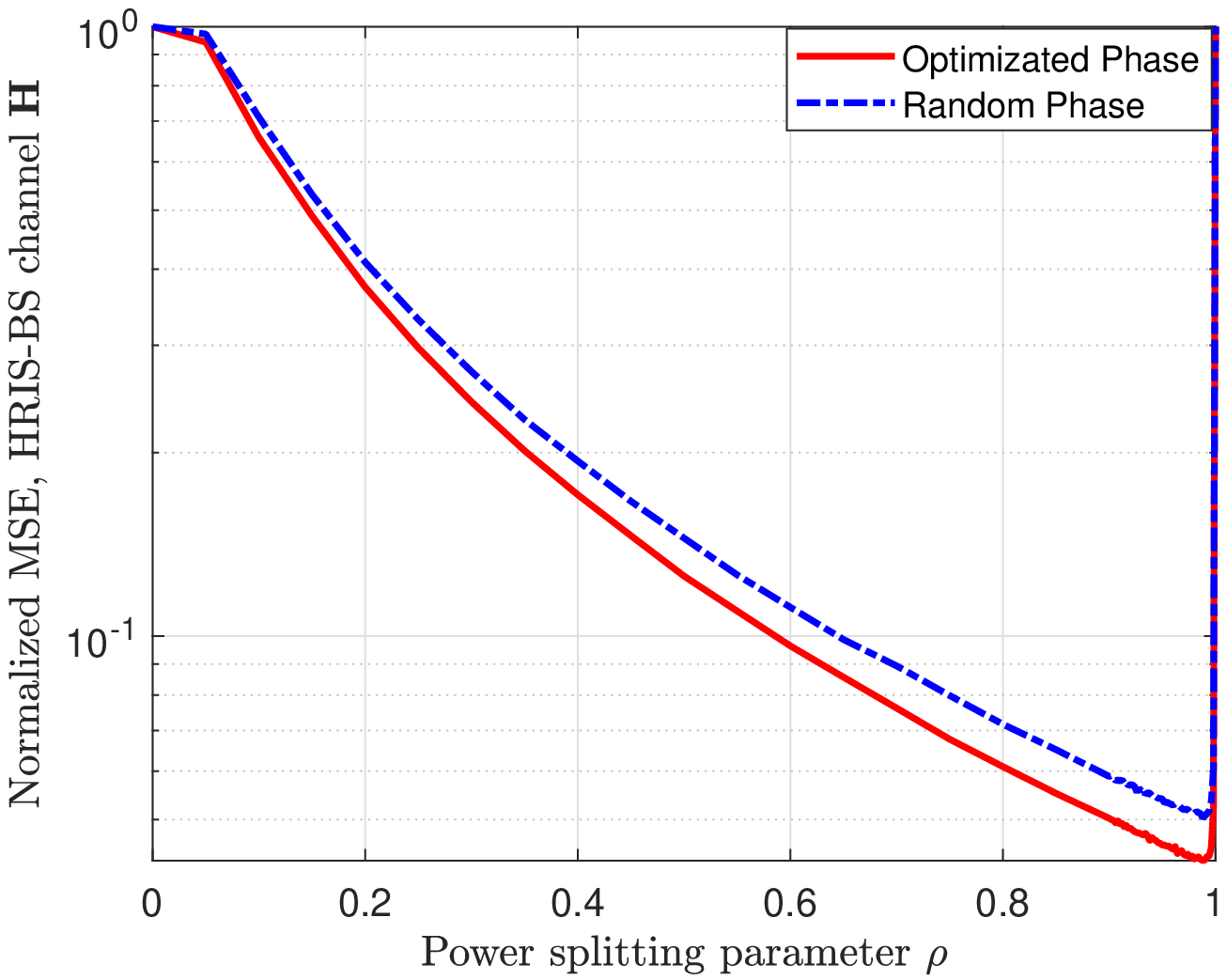} 
  } 
  \caption{Normalized \ac{mse} performance of the proposed channel estimation algorithm in recovering the combined \acp{ut}-HRIS channel ${\bf G}$ at the HRIS and the HRIS-\ac{bs} channel ${\bf H}$ at the \ac{bs} using $\tau=104$ pilots symbols.} 
  \label{fig:tradeoff} 
\end{figure}

In Fig.~\ref{fig:tradeoff}, we show the trade-off between the normalized \ac{mse} performances when estimating the \acp{ut}-\ac{hris} channel ${\bf G}$ at the \ac{hris} and the \ac{hris}-\ac{bs} channel  ${\bf H}$  at the \ac{bs} for different values of the power splitting parameter $\rho$; we have assumed that all the elements of the \ac{hris} have the same power splitting parameter $\rho$. In the figure, we have set the transmit SNR to $\Gamma = 100$ dB and the pilot length $\tau = 104$. In addition, ``Optimized Phase" denotes that the phase variables of the \ac{hris} have been optimized using Algorithm~\ref{alg:computational_graph_optimization}, and ``Random Phase" indicates that the \ac{hris} phase variables were generated randomly. It can be observed from Fig.~\ref{fig:subfig:G} that the normalized MSE of estimating the individual channel ${\bf G}$ increases significantly as $\rho$ increases. This happens because as $\rho$ increases, each meta-atoms element of the \ac{hris} splits less power of the impinging signal for channel estimation. It is also shown in Fig.~\ref{fig:subfig:H} that as $\rho$ increases, the normalized MSE of estimating the individual channel ${\bf H}$ decreases gradually. This improvement ceases when $\rho$ approaches $1$, implying that the \ac{hris} behaves as a purely passive \ac{ris} and the individual channels cannot be disentangled. This clearly demonstrates the fundamental trade-off between the accuracy in estimating each of the individual channels, which is dictated by the way that the \ac{hris} splits the power of the impinging signal. 
In addition, it is evident from the figure that the channel estimation performance of the ``Optimized Phase" case is slightly better than that of the ``Random Phase," especially when the power splitting parameter $\rho$ approaches its maximum value. This performance enhancement will become more significant when also optimizing the power splitting parameter $\rho$, as will be demonstrated in the sequel.

We now compare the channel estimation performance, considering our proposed \ac{hris}-based approach (labeled as ``Proposed HRIS"), the existing passive reflective \ac{ris}-based approach of \cite{wang2020channel} (labeled as ``Reflective RIS"), and two baseline schemes (labeled as ``Proposed HRIS with Partial Connection" and ``Proposed HRIS with Random Parameters"). The former baseline scheme adopts a partially-connected analog combiner at the \ac{hris} receiver, i.e., each antenna element just connects to only one receive RF chain, like the \ac{dma} structure defined in \cite{shlezinger2019dynamic}, and the corresponding power splitting parameters and phase configurations are optimized using Algorithm~\ref{alg:computational_graph_optimization}. With the latter baseline scheme, the power splitting parameters and phase profiles are randomly generated within the feasible set. In addition, since the channel estimation approach proposed in \cite{wang2020channel} can only estimate the cascaded channel, we also calculated the resulting cascaded channel estimation for the proposed approach, using our estimated individual BS-\ac{hris} channel $\myVec{\hat H}$  and the UTs-\ac{hris} channel $\myVec{\hat G}=[{\myVec{\hat g}}_1,{\myVec{\hat g}}_2,\ldots,{\myVec{\hat g}}_K]$. Specifically, the normalized MSE performance of the estimated cascaded channel with our \ac{hris}-based approach was calculated as follows:
\begin{align}
\operatorname{e}^{c} \triangleq \frac{\mathbb{E}\left\{\sum_{k=1}^{K}\left\|\myVec{\hat H}{\rm diag}\left({\myVec{\hat g}}_k\right)-\myVec{H}{\rm diag}\left({\myVec{ g}}_k\right)\right\|_{F}^{2}\right\}}{\mathbb{E}\left\{\sum_{k=1}^{K}\left\|\myVec{H}{\rm diag}\left({\myVec{ g}}_k\right)\right\|_{F}^{2}\right\}}.
	\label{eq:nmse}
\end{align}
\begin{figure}
    \centering
    \includegraphics[width=0.7\columnwidth]{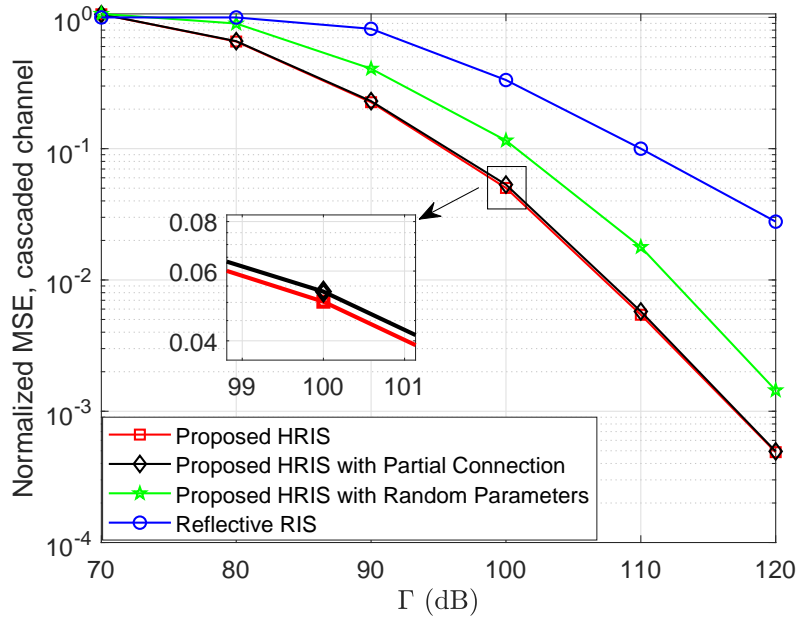}
    \caption{Normalized \ac{mse} performance of the cascaded channel estimation versus the transmit SNR $\Gamma$ for $\tau = 104$ pilots symbols for each of the $K=8$ UTs. Various versions of the proposed HRIS with $N_r=8$ receive RF chains and a reflective RIS have been considered.
    }
    \label{fig:comparison}
\end{figure}
\begin{figure}
    \centering
    \includegraphics[width=0.7\columnwidth]{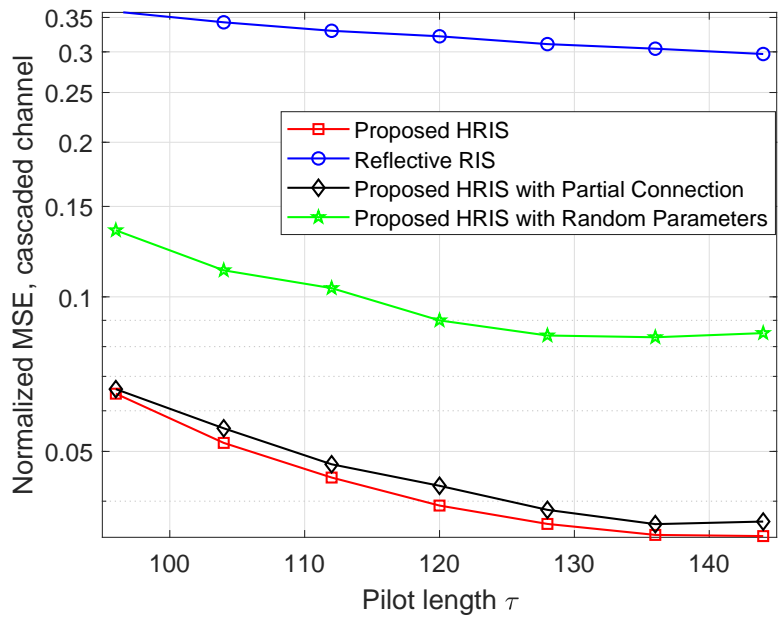}
    \caption{Normalized \ac{mse} performance of the cascaded channel estimation versus the pilot symbols' length $\tau$ used by each of the $K=8$ UTs, considering the transmit SNR $\Gamma = 100$~dB. All schemes compared in Fig.~\eqref{fig:comparison} have been considered.
    }
    \label{fig:pilot length}
\end{figure}

In Fig.~\ref{fig:comparison}, we demonstrate the normalized MSE of the cascaded channel estimation as a function of the transmit SNR value $\Gamma$ for the considered estimation methods, using $N_r = 8$ receive RF chains for our HRIS architecture. The method ``Reflective RIS'' in  \cite{wang2020channel} was found to require at least $92$ pilot symbols, and thus, we set the pilot length to $\tau=104$. As shown in the figure, the \acp{hris} sensing capability is translated into improved cascaded channel estimation accuracy, as compared to the state of the art. For example, the proposed \acp{hris} (even with the random configuration case ``Proposed HRIS with Random Parameters'') can achieve a much lower normalized MSE than that of the ``Relective RIS.'' In addition, it can be observed that the performance of the case ``Proposed HRIS with Partial Connection'' is comparable with that of the ``Proposed HRIS,'' which indicates the effectiveness of our proposed \ac{hris} even when using the lower power consumption and hardware complexity partially-connected analog combiner.

We also plot the normalized MSE of the cascaded channel against the pilot sequence length $\tau$ in Fig.~\ref{fig:pilot length}, considering the transmit SNR value $\Gamma=100$~dB and $N_r = 8$ received RF chains at the HRIS. Evidently, the proposed \ac{hris}-based approaches can significantly reduce the error of the cascaded channel estimation, as compared to the passive reflection \ac{ris}. Moreover, we see that for the proposed \ac{hris}, the normalized MSE decreases rapidly at first and then tends to slow down as the pilot length increases.
As before, our proposed HRIS with a partially-connected analog combiner is capable of achieving comparable channel estimation accuracy with that of HRIS with a fully-connected analog combiner.
\begin{figure}
    \centering
    \includegraphics[width=0.7\columnwidth]{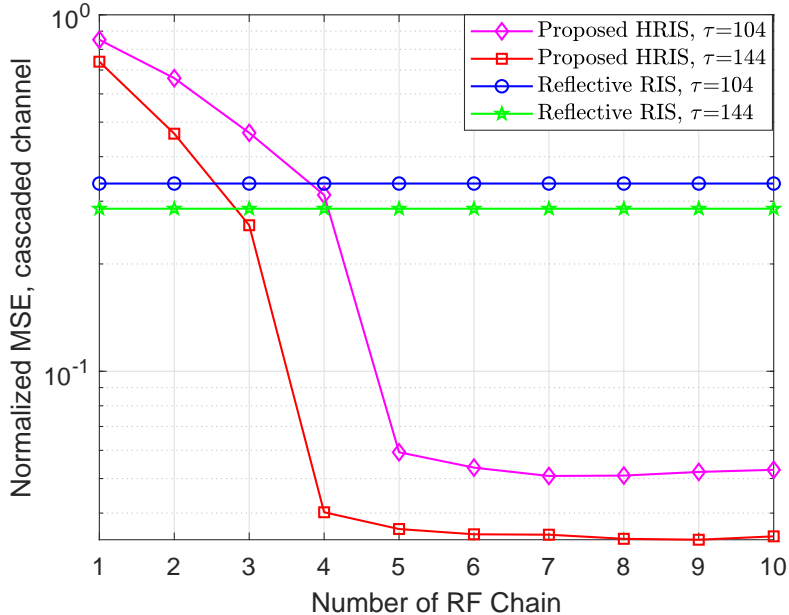}
    \caption{Normalized \ac{mse} performance of the cascaded channel estimation versus the number of the receive RF chains $N_r$ at the HRIS, considering the transmit SNR $\Gamma = 100$ dB and $\tau=\{104,144\}$ pilot symbols' length at each of the $K=8$ UTs. The proposed HRIS and a reflective RIS are compared.}
    \label{fig:RF-chain}
\end{figure}

Finally, in Fig.~\ref{fig:RF-chain}, we investigate the effect of the number $N_r$ of receive RF chains on the cascaded channel estimation accuracy, considering the same transmit SNR with Fig.~\ref{fig:pilot length} and different pilot sequence lengths $\tau$. It is illustrated that, as $N_r$ increases, the normalized cascaded channel MSE decreases initially and then converges to a constant value. This happens because the observed pilot signals increase proportionately to the number of receive RF chains, thereby increasing the accuracy of channel estimation. However, once the observed pilot signals exceed a threshold, increasing their number further will not reduce the channel estimation error. Moreover, it is evident from the figure that the proposed \ac{hris}-based approach achieves higher channel estimation accuracy than the one based on the conventional passive \ac{ris}, even when only few receive RF chains are used. For example, in the case of $\tau=104$, the proposed \ac{hris} requires only $N_r=5$ receive RF chains to achieve significantly improved channel estimation performance. Additionally, we see that the required $N_r$ value decreases as the pilot length increases.



\section{Conclusion}
\label{sec:Conclusions}
In this paper, we studied wireless communications aided by HRISs, which are metasurfaces capable of simultaneously reflecting and sensing impinging signals in a dynamically controllable manner. We presented a simple model for the operation of HRIS-empowered multi-user MIMO communications systems, and investigated their potential to facilitate channel estimation, as an indicative application. We showed that for the case without noise, HRISs enable to significantly save pilot overhead compared to that required by purely reflective \acp{ris}. We also quantified the achievable estimation error performance for the case with noise. In particular, we derived  the individual \ac{mse} for estimating individual channels at the \acp{hris} and \ac{bs}, and proposed a gradient-based approach to configure the \acp{hris} for minimizing the weighted sum-\ac{mse}.
 Our simulation results showcased the impactful role HRISs in RIS-empowered communications in estimating the individual channels as well as the cascaded channel over existing methods relying on nearly passive and reflective \acp{ris}.

\ifFullVersion
	%

	\begin{appendices}
	
		\numberwithin{proposition}{section} 
	\numberwithin{lemma}{section} 
	\numberwithin{corollary}{section} 
	\numberwithin{remark}{section} 
	\numberwithin{equation}{section}	
	

	%
	\vspace{-0.2cm}
	\section{Proof of Proposition \ref{pro:Noiseless}}
	\label{app:Proof1}	

By discarding the noise term in \eqref{eqn:LinearRC}, the received pilot signal at the \ac{hris} is given by
\begin{equation}
    \label{eqn:LinearRC_w/o}
    \myVec{y}_{\rm RC} = \myMat{A}_{\rm RC}\left( \left\{\myVec{\rho}(b), \myVec{\phi}(b)\right\}\right) {\myVec{G}} \myVec{S}.
\end{equation}
By using the identity $\operatorname{vec}(\mathbf{A B C})=\left(\mathbf{C}^{T} \otimes \mathbf{A}\right) \operatorname{vec}(\mathbf{B})$, we can rewrite \eqref{eqn:LinearRC_w/o} as follows:
\begin{equation}
    \label{eqn:LinearRC_without}
    \vecc\left(\myVec{y}_{\rm RC}\right) =\left( \myVec{S}^T \otimes \myMat{A}_{\rm RC}\left( \left\{\myVec{\rho}(b), \myVec{\phi}(b)\right\}\right) \right){\rm vec}(\myMat{G}).
\end{equation}
In the latter expression, we define $\myMat{A}_{\rm 1} \triangleq \myVec{S}^T \otimes \myMat{A}_{\rm RC}\left( \left\{\myVec{\rho}(b), \myVec{\phi}(b)\right\}\right)$. If $\myMat{A}_{\rm 1}$ is a full-column-rank matrix, then $\operatorname{vec}\left(\myMat{G}\right)$ can be recovered from \eqref{eqn:LinearRC_without} as
%
$\operatorname{vec}\left(\myMat{G}\right) = \myMat{A}_{\rm 1}^{\dagger} \vecc\left(\myVec{y}_{\rm RC}\right)$,
%
where $\myMat{A}_{\rm 1}^{\dagger}=\left(\myMat{A}_{\rm 1}^H \myMat{A}_{\rm 1}\right)^{-1}\myMat{A}_{\rm 1}^H$ is the pseudoinverse of $\myMat{A}_{\rm 1}$. Once we obtain the perfect estimate of $\operatorname{vec}\left(\myMat{G}\right)$, then the channel matrix ${\bf G}$ can be recovered accordingly. The dimension of $\myMat{A}_{\rm 1}$ is $N_r \tau$ by $N K$, and recall that $\tau=BT$. Thus, in order to guarantee that $\myMat{A}_{\rm 1}$ has a full column rank, the pilot length $\tau$ should satisfy the following inequality:
\begin{equation}\label{eq:pilot_length_1}
\tau \geq  \frac{NK}{N_r}.
\end{equation}

On the other hand, by discarding the noise term in \eqref{eq:y_BS}, the received pilot signal at the BS during $T$ time slots for each $b$-th sub-frame can be expressed as
%
\begin{equation}\label{eq:y_BS_w/o}
\myVec{y}_{\rm BS}\left( b \right) = \myVec{H} {\boldsymbol \Psi}\left( {\boldsymbol \rho}(b), {\boldsymbol \psi}(b)\right) \myVec{G} \myVec{S},
\end{equation}
or equivalently,
\begin{equation}\label{eq:y_BS_w/o}
\vecc\left(\myVec{y}_{\rm BS}\left( b \right)\right) =\left( \left( {\boldsymbol \Psi}\left( {\boldsymbol \rho}(b), {\boldsymbol \psi}(b)\right) \myVec{G} \myVec{S}\right)^T \otimes \myVec{I}_M \right) {\rm vec}(\myMat{H}).
\end{equation}
In the sequel, we make use of the notation $\myVec{\bar y}_{\rm BS}$ for the $M\tau\times 1$ vector generated by stacking the vectors $\vecc \left(\myVec{ y}_{\rm BS}\left( 1 \right)\right),\vecc \left(\myVec{y}_{\rm BS}\left( 2\right)\right),\ldots, \vecc \left(\myVec{y}_{\rm BS}\left( B \right)\right)$. It follows from \eqref{eq:y_BS_w/o} that we can express $\myVec{y}_{\rm BS}$ as
    $\myVec{\bar y}_{\rm BS} = \myMat{ A}_{\rm 2}\vecc \left(\myVec{H}\right)$,
where $\myMat{A}_{\rm 2} \in \mathbb{C}^{M\,\tau \times M\,N}$ is given by 
\begin{equation} \label{eq:Ab}
   \myMat{ A}_{\rm 2} = \left[ {\boldsymbol \Psi}\left( {\boldsymbol \rho}(1), {\boldsymbol \psi}(1)\right) \myVec{G} \myVec{S},\cdots, {\boldsymbol \Psi}\left( {\boldsymbol \rho}(B), {\boldsymbol \psi}(B)\right) \myVec{G} \myVec{S}\right]^T\otimes \myVec{I}_M,
\end{equation}
Similarly to $\myMat{A}_{\rm 1}$ before, we can perfectly recover ${\rm vec}(\myMat{H})$ if $\myMat{A}_{\rm 2}$ is a full-column-rank matrix, i.e., it holds:
%
${\rm vec}(\myMat{H}) =\myMat{A}_{\rm 2}^{\dagger} \myVec{\bar y}_{\rm BS}$,
%
where $\myMat{A}_{\rm 2}^{\dagger} =\left(\myMat{A}_{\rm 2}^H \myMat{A}_{\rm 2}\right)^{-1}\myMat{A}_{\rm 2}^H$. In order to guarantee that $\myMat{A}_{\rm 2}$ has a full column rank, the pilot length $\tau$ should satisfy the following inequality:
\begin{equation}\label{eq:pilot_length_2}
\tau \geq  N.
\end{equation}

By putting \eqref{eq:pilot_length_1} and \eqref{eq:pilot_length_2} together, we conclude that the number of pilots $\tau$ should satisfy the inequality:
\begin{equation}
    \label{eqn:Pilot_length}
    \tau \geq N \max\left\{1,\frac{K}{N_r}\right\},
\end{equation}
which completes the proof of Proposition \ref{pro:Noiseless}.

\vspace{-0.2cm}
\section{Proof of Theorem \ref{Theorem:1}}
\label{app:Proof2}

For notation brevity, we define $\myMat{\Phi} \triangleq \left\{\myVec{\rho}(b), \myVec{\phi}(b)\right\}$ during the proof of Theorem \ref{Theorem:1}. To estimate the channels between the HRIS and the UTs, we project $\myVec{y}_{\rm RC}$ defined in \eqref{eqn:LinearRC} on $\myVec{S}^H$, yielding
\begin{equation}\label{eq:estimate_g_k}
    \myVec{\tilde y}_{\rm RC}=\frac{1}{T} \myVec{y}_{\rm RC} \myVec{S}^H= \myMat{A}_{\rm RC}\left( \myMat{\Phi}\right) \myVec{G} + \myVec{\tilde z}_{\rm RC},
\end{equation}
where $\myVec{\tilde z}_{\rm RC} \triangleq \frac{1}{T}\myVec{z}_{\rm RC}\myVec{S}^H$, whose distribution is given by $C \mathcal{N}\left(0, K(T \Gamma)^{-1} \myVec{ I}_{N_r B}\right) $. According to \eqref{eq:estimate_g_k}, the linear estimation that minimizes the mean-square-error (MSE) of the estimation of $\myVec{G}$ has the following form \cite{biguesh2006training}:
\begin{equation}\label{eq:LMMSE_g_k}
   \myVec{\hat G}= \myVec{M}_0 \myVec{\tilde y}_{\rm RC},
\end{equation}
where $\myVec{M}_0$ is the linear estimator, which can be obtained by solving the following problem:
\begin{equation}\label{eq:MMSE_estimator}
\begin{split}
\myVec{M}_o & \triangleq\arg \min _{\myVec{M}} {\mathbb{E}} \left\{\left\|\myVec{G}-{\myVec{\hat G}}\right\|_{F}^{2} \right\}  = \arg \min _{{\myVec{M}}} {\mathbb{E}} \left\{\left\|\myVec{G} - \myVec{M} \myVec{\tilde y}_{\rm RC}\right\|_{F}^{2}\right\}.
\end{split}
\end{equation}
The error of this estimator can be given by computed as follows:
\begin{equation}\label{eq:MSE}
\begin{split}
\mathcal{ E}_{\myVec{G}} \left(\myMat{\Phi}\right)=&  \mathbb{E} \left\{\left\|\myVec{G} - \myVec{M} \myVec{\tilde y}_{\rm RC}\right\|_{F}^{2}\right\}  \\
=& {\rm Tr} \left(\myVec{R}_{\rm G}\right) - {\rm Tr} \left(\myVec{R}_{\rm G}\, \myMat{A}_{\rm RC}\left( \myMat{\Phi}\right)^H\, {\myVec{M}}^H \right) - {\rm Tr} \left({\myVec{M}}\, \myMat{A}_{\rm RC}\left( \myMat{\Phi}\right)\, \myVec{R}_{\rm G}\right) \\
&+ {\rm Tr} \left({\myVec{M}}\, \left( \myMat{A}_{\rm RC}\left( \myMat{\Phi}\right)\, \myVec{R}_{\rm G}\, \myMat{A}_{\rm RC}\left( \myMat{\Phi}\right)^H + K\, (T \Gamma)^{-1}\, {\myVec{ I}}_{N_r B} \right)\,{\myVec{M}}^H \right),
\end{split}
\end{equation}
where $ \myVec{R}_{\rm G} = \mathbb{E}\left[{\myVec{G}} {\myVec{G}}^H \right] = \left(\sum_{k=1}^{K} \gamma_k\right) \myVec{ I}_N$ denotes the covariance matrix of ${\myVec{G}}$. 
Since the second-order channel statistics vary slowly with time in general, here we assume that $ \myVec{R}_{\rm G}$ can be perfectly estimated at the HRIS. The optimal ${\myVec{M}}_o$ can be found from $\partial \mathcal{ E}_{\myVec{G} } \left(\myMat{\Phi}\right) / \partial \myVec{M} = 0$
and is given by
\begin{equation} \label{eq:optimal_MSE_G}
\begin{split}
 {\myVec{M}}_o =   \myVec{R}_{\rm G}\,  \myMat{A}_{\rm RC}\left( \myMat{\Phi}\right)^H\, \left( \myMat{A}_{\rm RC}\left( \myMat{\Phi}\right)\, \myVec{R}_{\rm G}\, \myMat{A}_{\rm RC}\left( \myMat{\Phi}\right)^H + K\, (T \Gamma)^{-1}\,{\myVec{I}_{N_r B}} \right)^{-1}.
 \end{split}
\end{equation}     
Substituting \eqref{eq:optimal_MSE_G} into \eqref{eq:MSE} and using $\left(\myVec{A}+\myVec{B C D}\right)^{-1}=\myVec{A}^{-1}-\myVec{A}^{-1} \myVec{B}\left(\myVec{D} \myVec{A}^{-1} \myVec{B}+ \myVec{C}^{-1} \right)^{-1} \myVec{D} \myVec{A}^{-1}$, the MMSE estimation error of $\myVec{G}$ can be derived as
\begin{equation}\label{eq:MMSE_error_G}
\mathcal{ E}_{\myVec{G}} \left(\myMat{\Phi}\right)  = {\rm Tr} \left\{\left(\myVec{R}_{\rm G}^{-1}\, +T \Gamma\, K^{-1}\myMat{A}_{\rm RC}\left( \myMat{\Phi}\right)^H\, \myMat{A}_{\rm RC}\left( \myMat{\Phi}\right)\right)^{-1}\right\}.
\end{equation}

The linear MMSE estimator of $\myMat{G}$ can be expressed as  
\begin{equation}\label{eq:LMMSE_G}
   {\myVec{\hat G}}= \myVec{R}_{\rm G}\,  \myMat{A}_{\rm RC}\left( \myMat{\Phi}\right)^H\, \left( \myMat{A}_{\rm RC}\left( \myMat{\Phi}\right)\, \myVec{R}_{\rm G}\, \myMat{A}_{\rm RC}\left( \myMat{\Phi}\right)^H + K\, (T \Gamma)^{-1}\,{\myVec{ I}}_{N_r B} \right)^{-1}\,\myVec{\tilde y}_{\rm RC},
\end{equation}
and it is easy to verify that the mean of ${\myVec{\hat G}}$ is zero, i.e., $ \mathbb{E} \left\{ {\myVec{\hat G}} \right\} =0$. Its covariance matrix is given by
\begin{equation}\label{eq:COvariance_G}
\begin{split}
{\boldsymbol \Sigma} \left(\myMat{\Phi}\right) & = \mathbb{E} \left\{ {\myVec{\hat G}}\, {\myVec{\hat G}}^H \right\}\\
& = \mathbb{E} \left\{ {\myVec{M}}_o \left(\myMat{A}_{\rm RC}\left( \myMat{\Phi}\right)\, \myVec{G}\, \myVec{G}^H\, \myMat{A}_{\rm RC}\left( \myMat{\Phi}\right)^H + {\bf Z}_r\right){\myVec{M}}_o^H \right\}\\
& = {\myVec{M}}_o \left(\mathbb{E} \left\{ \myMat{A}_{\rm RC}\left( \myMat{\Phi}\right)\, \myVec{G}\, \myVec{G}^H\, \myMat{A}_{\rm RC}\left( \myMat{\Phi}\right)^H \right\} +   K\,  (T \Gamma)^{-1}\,{\myVec{ I}}_{N_r B}\right){\myVec{M}}_o^H \\
& = {\myVec{M}}_o \left( \myMat{A}_{\rm RC}\left( \myMat{\Phi}\right)\, \myVec{R}_{\rm G}\, \myMat{A}_{\rm RC}\left( \myMat{\Phi}\right)^H  +  K\,  (T \Gamma)^{-1}\,{\myVec{ I}}_{N_r B}\right){\myVec{M}}_o^H \\
& = \myVec{R}_{\rm G}\,  \myMat{A}_{\rm RC}\left( \myMat{\Phi}\right)^H\, \left( \myMat{A}_{\rm RC}\left( \myMat{\Phi}\right)\, \myVec{R}_{\rm G}\, \myMat{A}_{\rm RC}\left( \myMat{\Phi}\right)^H + K\,  (T \Gamma)^{-1}\,{\myVec{ I}}_{N_r B} \right)^{-1} \myMat{A}_{\rm RC}\left( \myMat{\Phi}\right) \myVec{R}_{\rm G}^H.
\end{split}
\end{equation}
We finally let $\myVec{\tilde G} = \myVec{G} - \myVec{\hat G}$ denote the channel estimation error, which has zero mean and the covariance matrix 
\begin{equation}\label{eq:error_MSE}
\myVec{R}_{\bf \tilde G} \left(\myMat{\Phi}\right)  =  \mathbb{E} \left\{\myVec{\tilde G}\, \myVec{\tilde G}^H\right\}  \\
= \left(\myVec{R}_{\rm G}^{-1}\, + T \frac{\Gamma}{K}\,\myMat{A}_{\rm RC}\left( \myMat{\Phi}\right)^H\, \myMat{A}_{\rm RC}\left( \myMat{\Phi}\right)\right)^{-1}.
\end{equation}
The latter expression concludes the proof.
%



\vspace{-0.2cm}
\section{Proof of Theorem \ref{Theorem:2}}
\label{app:Proof3}

For notation brevity, we define $\myMat{\Phi} \triangleq \left\{\myVec{\rho}(b), \myVec{\phi}(b)\right\}$,  $\myMat{\Omega} \triangleq \left\{ \myVec{\rho}(b), \myVec{\phi}(b), \myVec{\psi}(b)\right\}$, and $\myMat{\Psi}(b) \triangleq \myMat{\Psi}(\myVec{\rho}(b),\myVec{\psi}(b))$, during the proof of Theorem \ref{Theorem:2}. By projecting $\myVec{y}_{\rm BS}\left( b \right)$ defined in \eqref{eq:y_BS} on $\myVec{S}^H$ and scaling the resulting term by $1/T$, we get the expression: 
\begin{equation}\label{eq:estimation_BS_project}
\myVec{\tilde y}_{\rm BS}\left( b \right) = \frac{1}{T} \myVec{y}_{\rm BS}[l]  \myVec{S}^H=  \myVec{H} {\boldsymbol \Psi}\left(b\right) \myVec{G}  + \myVec{\tilde z}_{\rm BS}\left( b \right),
\end{equation}
where $\myVec{\tilde z}_{\rm BS}\left( b \right)\triangleq\frac{1}{T} \myVec{ z}_{\rm BS}\left( b \right) \myVec{S}^H$.
By applying the identity $\operatorname{vec}(\mathbf{A B C})=\left(\mathbf{C}^{T} \otimes \mathbf{A}\right) \operatorname{vec}(\mathbf{B})$, we rewrite $\myVec{\tilde y}_{\rm BS}[l]$ in \eqref{eq:estimation_BS_project} in the following vector form:
\begin{equation}\label{eq:vector_form_error}
\begin{split}
& \vecc \left(\myVec{\tilde y}_{\rm BS}\left( b \right)\right)\\
&= \left(\myVec{G}^T {\boldsymbol \Psi}\left(b\right)^T \otimes \myVec{I}_{M}\right) \vecc \left(\myVec{H}\right) +  \vecc \left(\myVec{\tilde z}_{\rm BS}\left( b \right)\right)\\
&=\left((\myVec{\hat G} +\myVec{\tilde G})^T {\boldsymbol \Psi}\left( b\right)^T \otimes \myVec{I}_M\right) \vecc \left(\myVec{H}\right) +  \vecc \left(\myVec{\tilde z}_{\rm BS}\left( b \right)\right)\\
& = \left(\myVec{\hat G}^T {\boldsymbol \Psi}\left(b\right)^T \otimes \myVec{I}_M\right) \vecc \left(\myVec{H}\right) + \left(\myVec{\tilde G}^T {\boldsymbol \Psi}\left( b\right)^T \otimes \myVec{I}_M\right) \vecc \left(\myVec{H}\right)   +  \vecc \left(\myVec{\tilde z}_{\rm BS}\left( b \right)\right)
\end{split}
\end{equation}
where $\myVec{\tilde G} = \myVec{G}-\myVec{\hat G}$ denotes the channel estimation error of $\myVec{G}$.

Let the notation $\myVec{y}_{\rm BS}$ represent the $MKB\times 1$ vector generated by stacking the following vectors: $\vecc \left(\myVec{\tilde y}_{\rm BS}\left( 1 \right)\right),\vecc \left(\myVec{\tilde y}_{\rm BS}\left( 2\right)\right),\ldots, \vecc \left(\myVec{\tilde y}_{\rm BS}\left( b \right)\right)$. We can express $\myVec{y}_{\rm BS}$ from \eqref{eq:vector_form_error} as
\begin{equation}
    \label{eqn:LinearBS}
    \myVec{y}_{\rm BS} = \myMat{A}_{\rm BS} {\myVec{h}}   +  \underbrace{\Delta \myMat{A}_{\rm BS} {\myVec{h}} + \myVec{z}_{\rm BS} }_{\myVec{z}},
\end{equation}
where $\myVec{z}_{\rm BS}$ results from stacking $\vecc \left(\myVec{\tilde z}_{\rm BS}\left( 1 \right)\right), \vecc \left(\myVec{\tilde z}_{\rm BS}\left( 2\right)\right),\ldots,\vecc \left(\myVec{\tilde z}_{\rm BS}\left( B \right)\right)$ and $\myVec{h}=\vecc \left(\myVec{H}\right)$, as well as $\myMat{A}_{\rm BS} \in \mathbb{C}^{M\,K\, B\times M\,N}$ and $\Delta \myMat{A}_{\rm BS} \in \mathbb{C}^{M\,K\, B\times M\,N}$ are respectively given by 
\begin{equation} \label{eq:Ab}
   \myMat{A}_{\rm BS} \triangleq \left[\myMat{\Psi}(1) \myVec{\hat G},\cdots,\myMat{\Psi}(B) \myVec{\hat G}\right]^T\otimes \myVec{I}_M,
\end{equation}
\begin{equation} \label{eq:Delta_Ab}
  \Delta \myMat{A}_{\rm BS} \triangleq \left[\myMat{\Psi}(1) \myVec{\tilde G},\cdots,\myMat{\Psi}(B)  \myVec{\tilde G}\right]^T\otimes \myVec{I}_M
\end{equation}

We next define $\myVec{z} \triangleq \Delta \myMat{A}_{\rm BS} {\myVec{h}} + \myVec{z}_{\rm BS}$ as the effective noise vector at the BS, which includes the colored interference forwarded from the HRIS and the local AWGN vector  $\myVec{z}_{\rm BS}$. We present the following Proposition, which provides the second-order statistics of $\myVec{z}$.
\begin{proposition} 
\label{Prop:noise_covariance}
The covariance matrix of $\myVec{z}$ is given by $\myVec{R}_{z}\left(\myMat{\Omega} \right) = \mathbb{E}\left\{\myVec{z} \myVec{z}^H\right\} = {\bf D}\left(\myMat{\Omega} \right) \otimes {\bf I}_M$,
where $\myVec{D}\left(\myMat{\Omega} \right)$ is a ${BK \times BK}$ matrix, which can be partitioned into $B \times B$ blocks with each block being a $K \times K$ submatrix. The $i$-th row and $j$-th column block of $\myVec{D}\left(\myMat{\Omega} \right)$ is defined as
\begin{equation}\label{eq:D_Omega}
\begin{split}
&\left[\myVec{D}\left(\myMat{\Omega} \right)\right]_{i,j} \triangleq \left\{\begin{array}{ll}
\frac{\beta}{K} {\rm Tr}\left(\myMat{\Psi}(j)^H\myMat{\Psi}(i)\myVec{R}_{\bf \tilde G} \left(\myMat{\Phi}\right)\right)\myVec{I}_K + (T \Gamma)^{-1} \myVec{I}_{ K}, & {\rm if}~ i = j \\
\frac{\beta}{K} {\rm Tr}\left(\myMat{\Psi}(j)^H\myMat{\Psi}(i)\myVec{R}_{\bf \tilde G} \left(\myMat{\Phi}\right)\right)\myVec{I}_K, & {\rm if}~ i \neq j
\end{array}\right..
\end{split}
\end{equation}
\end{proposition}
\begin{IEEEproof}
We start with the definition of the covariance matrix of $\myVec{z}$:
\begin{equation} \label{eq:noise_cov1}
\myVec{R}_z\left(\myMat{\Omega} \right) = \mathbb{E}\left\{\myVec{z} \myVec{z}^H\right\}
 =  \mathbb{E}_{\myVec{\tilde G}, \myVec{h}}\left\{ \Delta \myMat{A}_{\rm BS} {\myVec{h}} {\myVec{h}}^H \Delta \myMat{A}_{\rm BS}^H  \right\} + \mathbb{E}_{{\bf  z}_b}\left\{{\bf  z}_b {\bf  z}_b^H\right\}.
\end{equation}
Since $\vecc \left(\myVec{\tilde z}_{\rm BS}\left( b \right)\right) = \frac{1}{T} \left(\myVec{S}^{\dag} \otimes  \myVec{I}_M\right) \vecc\left(\myVec{z}_{\rm BS}\left( b \right)\right)$, we can first calculate $\mathbb{E}_{\myVec{z}_{\rm BS}}\left\{\myVec{z}_{\rm BS} \myVec{ z}_{\rm BS}^H\right\}$ first as:
\begin{equation}\label{eq:noise_cov2}
\begin{split}
\mathbb{E}_{\myVec{z}_{\rm BS}}\left\{\myVec{z}_{\rm BS} \myVec{ z}_{\rm BS}^H\right\} &=\frac{1}{T^2} \mathbb{E}_{\myVec{z}_{\rm BS}}\left\{\left[\begin{array}{l}
\left(\myVec{S}^{\dag} \otimes \myVec{I}_M \right)\vecc\left(\myVec{z}_{\rm BS}\left( 1 \right)\right)\\
~~~~~\vdots \\
\left(\myVec{S}^{\dag} \otimes \myVec{I}_M \right)\vecc\left(\myVec{z}_{\rm BS}\left( B \right)\right)
\end{array}\right] \left[\begin{array}{l}
\left(\myVec{S}^{\dag} \otimes \myVec{I}_M \right)\vecc\left(\myVec{z}_{\rm BS}\left( 1 \right)\right)\\
~~~~~\vdots \\
\left(\myVec{S}^{\dag} \otimes \myVec{I}_M \right)\vecc\left(\myVec{z}_{\rm BS}\left( B \right)\right)
\end{array}\right]^H \right\}\\
& = (T \Gamma)^{-1}\myVec{I}_{B K} \otimes \myVec{I}_M
\end{split}
\end{equation}

On the other hand, the following expression can be deduced:
\begin{equation} \label{eq:noise_cov}
\begin{split}
& \mathbb{E}_{\myVec{\tilde G}, \myVec{h}}\left\{ \Delta \myMat{A}_{\rm BS} {\myVec{h}} {\myVec{h}}^H \Delta \myMat{A}_{\rm BS}^H  \right\} \\
&  \overset{(a)}{=} \beta \mathbb{E}_{\myVec{\tilde G}}\left\{\Delta \myMat{A}_{\rm BS} \Delta \myMat{A}_{\rm BS}^H\right\} \\
&= \beta \mathbb{E}_{\myVec{\tilde G}}\left\{\left[\begin{array}{lll}
{\myVec{\tilde G}}^H \myMat{\Psi}^H(1)\myMat{\Psi}(1)\,{\myVec{\tilde G}}&\cdots&{\myVec{\tilde G}}^H \myMat{\Psi}^H(1)\myMat{\Psi}(B)\,{\myVec{\tilde G}} \\
~~~~~\vdots & \ddots & ~~~~~\vdots \\
{\myVec{\tilde G}}^H \myMat{\Psi}^H(B)\myMat{\Psi}(1)\,{\myVec{\tilde G}} &\cdots&{\myVec{\tilde G}}^H \myMat{\Psi}^H(B)\myMat{\Psi}(B)\,{\myVec{\tilde G}}
\end{array}\right]^T \otimes \myVec{I}_{M}\right\}  \\
&\overset{(b)}{=}
\frac{\beta}{K} \left[\begin{array}{lll}
{\rm Tr}\left(\myMat{\Psi}^H(1)\myMat{\Psi}(1)\myVec{R}_{\bf \tilde G} \left(\myMat{\Phi}\right)\right)\myVec{I}_K & \cdots & {\rm Tr}\left(\myMat{\Psi}^H(1)\myMat{\Psi}(B)\myVec{R}_{\bf \tilde G} \left(\myMat{\Phi}\right)\right)\myVec{I}_K \\
~~~~~\vdots & \ddots & ~~~~~\vdots \\
{\rm Tr}\left(\myMat{\Psi}^H(B)\myMat{\Psi}(1)\myVec{R}_{\bf \tilde G} \left(\myMat{\Phi}\right)\right)\myVec{I}_K & \cdots &{\rm Tr}\left(\myMat{\Psi}^H(B)\myMat{\Psi}(B)\myVec{R}_{\bf \tilde G} \left(\myMat{\Phi}\right)\right)\myVec{I}_K
\end{array}\right]^T \otimes \myVec{I}_{M},
\end{split}
\end{equation}
where $(a)$ holds due to the fact that $\mathbb{E}_{\myVec{h}} \left\{\myVec{h} \myVec{h}^H\right\} = \beta {\bf I}_{MN}$ and $(b)$ results from the following derivation: %
\begin{equation}
\begin{split}
\mathbb{E}_{\myVec{\tilde G}}\left\{{\myVec{\tilde G}}^H \myMat{\Psi}(i)^H \myMat{\Psi}(j) {\myVec{\tilde G}}\right\}
\overset{(c)}{=}&\mathbb{E}_{\myVec{U}}\left\{\myVec{U}^H \myVec{R}_{\bf \tilde G} \left(\myMat{\Phi}\right)^{1/2}\myMat{\Psi}(i)^H \myMat{\Psi}(j) \myVec{R}_{\bf \tilde G} \left(\myMat{\Phi}\right)^{1/2} \myVec{U}\right\}\\
&= \frac{1}{K} {\rm Tr} \left( \myVec{R}_{\bf \tilde G} \left(\myMat{\Phi}\right)^{1/2}\myMat{\Psi}(i)^H \myMat{\Psi}(j) \myVec{R}_{\bf \tilde G} \left(\myMat{\Phi}\right)^{1/2}\right) \myVec{I}_K\\
&= \frac{1}{K} {\rm Tr} \left( \myMat{\Psi}(i)^H \myMat{\Psi}(j) \myVec{R}_{\bf \tilde G} \left(\myMat{\Phi}\right)\right) \myVec{I}_K.
\end{split}
\end{equation}
In the latter expression, $(c)$ is due to the fact that $\myVec{\tilde G}=\myVec{R}_{\bf \tilde G} \left(\myMat{\Phi}\right)^{1/2} \myVec{U}$ with $\myVec{U}\in \mathbb{C}^{N \times K}$ denoting a matrix satisfying $\myVec{U} \sim \mathcal{C} \mathcal{N}\left(\mathbf{0}, \myVec{I}_{N}\right)$.
By substituting \eqref{eq:noise_cov2} and \eqref{eq:noise_cov} into  \eqref{eq:noise_cov1} yields:
\begin{equation}
\myVec{R}_z\left(\myMat{\Omega} \right) = \myVec{D}\left(\myMat{\Omega} \right) \otimes \myVec{I}_M,
\end{equation}
where $\myVec{D}\left(\myMat{\Omega} \right)$ is the ${BK \times BK}$ matrix in \eqref{eq:D_Omega}, which can be partitioned into $B \times B$ blocks with each block being a $K \times K$ submatrix. 
%

\end{IEEEproof}

For notation brevity, in the following we make use of the simplified notations $\myMat{R}_{z}$ and $\myVec{D}$ to represent $\myVec{R}_{z}\left(\myMat{\Omega} \right)$ and $\myVec{D}\left(\myMat{\Omega} \right)$, respectively. We also utilize the LMMSE estimator to estimate $\myVec{h}$ from $\myVec{y}_{\rm BS}$ as $ \myVec{\hat h} = \myVec{T} \myVec{y}_{\rm BS}$, where $\myVec{T}$ is the optimal solution that minimizes the following channel estimation error:
\begin{equation}\label{eq:Vec_MSE}
\begin{split}
\mathcal{E}_{\myVec{h}}= \mathbb{E} \left\{\left\|\myVec{h} -\myVec{\hat h}  \right\|^{2}\right\}  = \mathbb{E} \left\{\left\|\myVec{h} -\myVec{T} \myVec{y}_{\rm BS} \right\|^{2}\right\}.
\end{split}
\end{equation}

It is well-known \cite{sengijpta1995fundamentals} that the the optimal $\myVec{T}$ can be obtained as 
\begin{equation}\label{eq:Vec_MSE_T}
\begin{split}
\myVec{T} &=\mathbb{E}\left[\myVec{h} \myVec{y}_{\rm BS}^H\right]\left(\mathbb{E}\left[\myVec{y}_{\rm BS} \myVec{y}_{\rm BS}^H\right]\right)^{-1} = \myVec{R}_h\, \myVec{A}_{BS}^H\, \left(\myVec{A}_{BS}\, \myVec{R}_h\, \myVec{A}_{BS}^H\, + \myVec{R}_{z} \right)^{-1},
\end{split}
\end{equation}
where $\myVec{R}_h = \mathbb{E}\left\{\myVec{h}\, \myVec{h}^H\right\} = \beta \myVec{I}_{MN}$. By substituting \eqref{eq:Vec_MSE_T} into \eqref{eq:Vec_MSE}, we get the expression: 
\begin{align}
\mathcal{E}_{\myVec{h}} &= \mathbb{E}_{\myVec{h},\myVec{\hat G}} \left\{\left\|\myVec{h} - \myVec{R}_h\, \myVec{A}_{BS}^H\, \left(\myVec{A}_{BS}\, \myVec{R}_h\, \myVec{A}_{BS}^H\, + \myVec{R}_z \right)^{-1} \myVec{y}_{\rm BS} \right\|^{2}\right\} \notag\\
 &=\mathbb{E}_{\myVec{\hat G}}\left\{ {\rm Tr} \left(\myVec{R}_h - \myVec{ R}_h\, \myVec{A}_{BS}^H\,  \left(\myVec{A}_{BS}\, \myVec{R}_h\, \myVec{A}_{BS}^H\, + \myVec{R}_z \right)^{-1}\, \myVec{A}_{BS}\, \myVec{R}_h\right) \right\} \notag\\
 & = \mathbb{E}_{\myVec{\hat G}}\left\{{\rm Tr} \left(\left(\myVec{R}_h^{-1} + \myVec{A}_{BS}^H\, \myVec{R}_z^{-1}\, \myVec{A}_{BS}\right)^{-1}\right)\right\} \notag\\
 &= \mathbb{E}_{\myVec{\hat G}}\left\{{\rm Tr} \left(\left(\myVec{R}_h^{-1} + \left[\myMat{\Psi}(1) \myVec{\hat G},\cdots,\myMat{\Psi}(B) \myVec{\hat G}\right]^{\dag}\myVec{D}^{-1}\left[\myMat{\Psi}(1) \myVec{\hat G},\cdots,\myMat{\Psi}(B) \myVec{\hat G}\right]^T \otimes \myVec{I}_M\right)^{-1}\right)\right\} \notag\\
&= \mathbb{E}_{\myVec{\hat G}}\left\{{\rm Tr} \left(\left(\myVec{R}_h^{-1} + \left( \sum_{j=1}^B \sum_{i=1}^B\,\myMat{\Psi}(i)\myVec{\hat G} \left[\myVec{D}^{-T}\right]_{i,j} \myVec{\hat G}^H \myMat{\Psi}(j)^H \right)^T
\otimes \myVec{I}_M\right)^{-1}\right)\right\}.
\label{eq:Vec_MSE_G}
\end{align}

Moreover, the following expression holds:
\begin{align}
\mathbb{E}_{\myVec{\hat G}}\left\{\myVec{\hat G} \left[\myVec{D}^{-T}\right]_{i,j} \myVec{\hat G}^H\right\}
&=\mathbb{E}_{\myVec{N}}\left\{{\boldsymbol \Sigma} \left(\myMat{\Phi}\right)^{1/2}\myVec{N} \left[\myVec{D}^{-T}\right]_{i,j} \myVec{N}^H {\boldsymbol \Sigma} \left(\myMat{\Phi}\right)^{1/2}\right\}\notag \\
&=K\, {\rm Tr}\left(\left[\myVec{D}^{-T}\right]_{i,j}\right) {\boldsymbol \Sigma} \left(\myMat{\Phi}\right),
\label{eq:temp1}
\end{align}
where ${\myVec{N}} \in \mathbb{C}^{N \times K}$ denotes a random matrix distributed as $\myVec{N} \sim \mathcal{C} \mathcal{N}\left(\mathbf{0},  \myVec{I}_{N}\right)$. In order to efficiently design the reflection and reception weights of the proposed HRIS, we next approximate the MSE defined in \eqref{eq:Vec_MSE_G} with a deterministic MSE, using a standard bounding technique. In particular, we lower-bound the MSE as follows:
\begin{align*}
\mathcal{E}_{\myVec{h}} & = \mathbb{E}_{\myVec{\hat G}}\left\{{\rm Tr} \left(\left(\myVec{R}_h^{-1} + \left( \sum_{j=1}^B \sum_{i=1}^B\,\myMat{\Psi}(i)\myVec{\hat G} \left[\myVec{D}^{-T}\right]_{i,j} \myVec{\hat G}^H \myMat{\Psi}(j)^H \right)^T
\otimes \myVec{I}_M\right)^{-1}\right)\right\}\\
& \overset{(a)}{\geq}   {\rm Tr} \left(\left(\myVec{R}_h^{-1} + \left( \sum_{j=1}^B \sum_{i=1}^B\,\myMat{\Psi}(i)\mathbb{E}_{\myVec{\hat G}}\left\{\myVec{\hat G} \left[\myVec{D}^{-T}\right]_{i,j} \myVec{\hat G}^H \right\} \myMat{\Psi}(j)^H \right)^T
\otimes \myVec{I}_M\right)^{-1}\right)\\
& \overset{(b)}{=}   {\rm Tr} \left(\left(\myVec{R}_h^{-1} + \left(K \sum_{j=1}^B \sum_{i=1}^B\,{\rm Tr}\left(\left[\myVec{D}^{-T}\right]_{i,j}\right)\, \myMat{\Psi}(i){\boldsymbol \Sigma} \left(\myMat{\Phi}\right) \myMat{\Psi}(j)^H \right)^T
\otimes \myVec{I}_M\right)^{-1}\right),
\end{align*}
where $(a)$ holds from the Jensen’s inequality and the fact that $\rm{Tr}\left(\mathbf{X}^{-1}\right)$ is a convex function with respect to $\mathbf{X}$. In addition, $(b)$ comes from \eqref{eq:temp1}. 

Putting all above together, we obtain the following lower bound of the channel estimation error at the BS: 
\begin{equation*}
\mathcal{E}_{\myVec{H}}\left( \myMat{\Omega} \right)= {\rm Tr} \left(\left(\myVec{R}_h^{-1} + \left(K \sum_{j=1}^B \sum_{i=1}^B\,{\rm Tr}\left(\left[\myVec{D}^{-T}\right]_{i,j}\right)\, \myMat{\Psi}(i){\boldsymbol \Sigma} \left(\myMat{\Phi}\right) \myMat{\Psi}(j)^H \right)^T
\otimes \myVec{I}_M\right)^{-1}\right),
\end{equation*}
which concludes the proof.

\end{appendices}	
\fi 

\vspace{-0.3cm}
\bibliographystyle{IEEEtran}
\bibliography{IEEEabrv,refs}

\end{document}